\documentclass{aa}
\usepackage{pdflscape}
\usepackage{soul}
\usepackage{color}
\usepackage{times}
\usepackage{natbib}
\usepackage{amsmath,amssymb,amsfonts,txfonts}
\usepackage[colorinlistoftodos]{todonotes}
\usepackage{multirow}
\usepackage{cellspace, hhline}
\usepackage{nicefrac}
\usepackage{notes2bib}
\usepackage{hyperref}   
\hypersetup{colorlinks=true,linkcolor=[rgb]{1.,0.2,0.2},citecolor=[rgb]{0.1,0.4,1.},filecolor=[rgb]{0.7,0.2,0.2},urlcolor=[rgb]{0.7,0.2,0.2}}
\newcommand {\vel}{$\,{\rm km}\,{\rm s}^{-1}$}
\newcommand {\ca}{$\rm C\mbox{-}17$}
\definecolor{darkspringgreen}{rgb}{0.09, 0.45, 0.27}

\definecolor{uva}{rgb}{0.45, 0.09, 0.45}
\newcommand {\Sec}{Sec.\,}

\newcommand {\Eq}{Eq.\,}
\usepackage{graphicx}
\usepackage{txfonts}

\begin{document} 

   \title{Cluster Strong Lensing with Hierarchical Inference}

   \subtitle{Formalism, functional tests, and public code release}

\author{
   P. Bergamini \inst{\ref{inafbo}} \and
   A. Agnello\inst{\ref{dark}} \and
   G.~B.~Caminha \inst{\ref{Kapteyn},\ref{MPA}}  
          }
   \authorrunning{Bergamini, Agnello, Caminha}

\institute{
    INAF - Osservatorio di Astrofisica e Scienza dello Spazio di Bologna, via Gobetti 93/3, I-40129 Bologna, Italy \label{inafbo}
   \and
   DARK, Niels Bohr Institute, University of Copenhagen,
    Jagtvej 128, 2200 Copenhagen \O, Denmark  \\\email{adriano.agnello@nbi.ku.dk} \label{dark}
   \and 
   Kapteyn Astronomical Institute, University of Groningen, Postbus 800, 9700 AV Groningen, The Netherlands \label{Kapteyn}
   \and 
   Max-Planck-Institut für Astrophysik, Karl-Schwarzschild-Str. 1, D-85748 Garching, Germany \label{MPA}
   }


 
  \abstract
   {Lensing by galaxy clusters is a versatile probe of cosmology and extra-galactic astrophysics, but the accuracy of some of its predictions is limited by the simplified models adopted to reduce the (otherwise intractable) number of degrees of freedom.}
   {We aim at cluster lensing models where the parameters of all cluster member galaxies are free to vary around some common scaling relations with non-zero scatter and deviate significantly from them if and only if the data require it.}
   {We have devised a Bayesian hierarchical inference framework that enables the determination of all lensing parameters and the scaling relation hyperparameters, including intrinsic scatter, from lensing constraints and (if given) stellar kinematic measurements. We achieve this through \textsc{BayesLens}, a purpose-built wrapper around common parametric lensing codes that can sample the full posterior on parameters and hyperparameters, which we release with this paper.}
   {We have run functional tests of our code against simple mock cluster lensing data-sets with realistic uncertainties. The parameters and hyperparameters are recovered within their 68\% credibility ranges and the positions of all the ``observed'' multiple images are accurately reproduced by the \textsc{BayeLens} best-fit model, without over-fitting.}
   {We have shown that an accurate description of cluster member galaxies is attainable, despite a large number of degrees of freedom, through fast and tractable inference. This extends beyond the state-of-the-art of current cluster lensing models. The precise impact on studies of cosmography, galaxy evolution, and high-redshift galaxy populations can then be quantified on real galaxy clusters. While other sources of systematics exist and may be significant in real clusters, our results show that the contribution of intrinsic scatter in cluster member populations can now be controlled.}


   \maketitle
%

\section{Introduction}
Galaxy clusters (at $z\approx0.3-0.9$) are ideal ``cosmic telescopes'' to study background galaxies out to $z\approx7$ and they boost the lensing signal of their own galaxies. When galaxies reside in clusters, their lensing cross-section is enhanced, allowing the study of galaxy populations at $z\approx0.2\mbox{-}\,0.4$, over a wide mass range (down to $M_{\star}\approx 10^{9.5} M_{\odot}$, e.g., \citealt{Keeton_2003,Grillo_2014,par16,nie17,ber19}). If the total mass density distributions of lensing clusters are accurately reconstructed, cosmological parameters can also be inferred \citep[e.g.,][]{gol02,gil09,cam16,mag18,gri18}. Dedicated Hubble Space Telescope (HST) surveys, such as the Cluster Lensing And Supernova survey with Hubble \citep[CLASH,][]{Postman_2012_clash}, the Hubble/Spitzer Frontier Fields program \citep[HFF,][]{lot17}, and the Re-ionization Lensing Cluster Survey \citep[RELICS,][]{Coe_2019}, have enabled the identification and multi-band characterization of tens, sometimes hundreds, of multiple images of background sources, and hundreds of cluster member galaxies per cluster so as to constrain cluster lens models \citep{pri17}. Subsequent multi-slit spectroscopic follow-up observations, such as the CLASH-VLT survey \citep{ros14} has gathered a wealth of multi-object spectroscopic data on cluster member galaxies and background sources, sampling light-cones around 13 clusters ($\approx20\times20$~arcmin$^2$). 
Due to their extent and magnifying power, galaxy clusters have also been used to study faint and high-redshift sources \citep[e.g.,][]{ishi15,liv15,kaw16,bou17,atek18,cer18,hoa18,van17,hashi19,van19}. Since these studies may have implications on our understanding of star-formation at high-redshift and re-ionization, accurate magnification maps are required.

The advent of the Multi-Unit Spectroscopic Explorer \citep[MUSE,][]{Bacon_MUSE} has enabled integral-field spectroscopy of galaxy clusters. In particular, in recent years numerous MUSE observations have targeted the cores of several massive clusters  \citep[e.g.,][]{Richard_2014,Jauzac_2015,Limousin_2016,Kawamata_2016,Lagattuta_2017,cam16,Caminha_macs0416,Caminha_macs1206,van19}.
Every MUSE pointing provides a data-cube with a field of view of 1 arcmin$^2$ and spatial sampling of 0.2\arcsec. The spectra cover ${4750<\lambda<9350}$\,\AA\ with a resolution of $\sim2.6$\,\AA,  almost constant along the whole wavelength range, and a spectral sampling of 1.25\,\AA/pix (the spectral resolution is robustly characterized, see \citealt{gue17}). Integral-field spectroscopy of cluster cores enabled a secure identification of tens of multiple images and their redshifts and complete and pure sets of member galaxies.

Besides the identification of cluster members and multiple images, spectroscopy provides further constraints on mass models through kinematics. The homogeneous spectroscopic coverage of CLASH-VLT means that velocities of individual cluster members can be used to aid the cluster profile reconstruction through dynamical modeling \citep{Biviano2013,Sartoris_2020}. Similarly, integral-field follow-up with MUSE yields the internal stellar kinematics of cluster members and allows to independently constrain their mass profiles.

Lensing clusters have undergone an extensive modeling effort by multiple independent teams. Up to now, the constraints have consisted in the positions of tens of multiple images per cluster \citep[e.g.,][]{Richard_2014,Grillo_2015, Jauzac_2015, Limousin_2016, Kawamata_2016, cam16, Lagattuta_2017, Caminha_macs0416, Caminha_macs1206, Bonamigo_2018, Caminha_2019, Bergamini_2020}. Mostly, lens models describe these clusters as a superposition of extended dark matter (DM) halos and more localized over-densities corresponding to cluster member galaxies. So-called ``non-parametric'', grid-based models \citep[e.g.,][]{dye98,saha01,bra05,die05} add mass over-densities only where they are required by the data. Alternatively, parametric models decompose the cluster potential in different (smooth) components that are expected to follow a physical hypothesis on how mass should be distributed in galaxies and clusters \citep[see, e.g.,][for an overview]{Jullo_lenstool}. In particular, \citet{Natarajan_1997} have argued that including cluster member galaxies explicitly, like individual components of cluster mass models, is crucial. This has also emerged from the HFF model comparison project \citep{Meneghetti_2017}, where parametric models routinely outperformed current versions of free-form models, both on semi-analytical and on fully simulated mocks. Towards the era of precision cosmology with galaxy clusters, free-form models may eventually replace parametric models, possibly resolving some of their built-in rigidity \citep[e.g.,][]{Rodney2018}. However, given the current performance of parametric and free-form models, one may safely adopt the working hypothesis that galaxy clusters can be described as a superposition of relatively simple galaxy/halo components.

The bottle-neck in cluster lens models consists in: $\approx0.3\arcsec \mbox{-}\,0.6\arcsec$ errors in image-position reconstruction \citep[e.g.,][]{Grillo_2016}, significantly higher than current positional uncertainties from HST data, and the appreciable discrepancies in magnification maps produced by different models on the same clusters \citep{Meneghetti_2017}. One cause of such discrepancies may be the rigidity of the models used to reduce the degrees of freedom associated with cluster members, since galaxies are currently modeled as belonging to zero-scatter scaling relations, typically with hyperparameters that are imposed externally \citep[e.g.,][]{Limousin_2007_truncation}.
The freedom and accuracy in lensing models have gained importance in the era of high-accuracy cosmography \citep[see, e.g.,][]{tre16}, and on the reliability of magnification maps for the study of high-redshift galaxy populations \citep{bou17}.
Up to now, departures of individual galaxies from prescribed scaling relations were determined heuristically, on a galaxy-by-galaxy basis \citep{jau18}. Ideally, all galaxies should be let free to vary around some finite-scatter scaling relations, whose parameters (including scatter) are to be determined directly from the data on each given cluster.

The solution is a hierarchical Bayesian inference formalism, where each galaxy has its own associated parameters, and the parameters of all galaxies are posited to be drawn from common relations with hyperparameters to be determined through lens modeling and (if given) auxiliary kinematic information. In this paper, we illustrate this formalism and its application to lensing models of three simple mock clusters of increasing complexity with HFF/CLASH-VLT data quality, accounting for all observational constraints in a self-consistent manner. This ensures that cluster lens models are as flexible as possible, given the data, and that higher accuracy is reached in predicted image positions. Since only one of the contributions to the inference is from lensing, and in order to ensure a fair comparison with state-of-the-art technology, we build our inference as a modular wrapper of common lens modeling codes. This also ensures that further constraints, such as time delays, flux-ratios, and shapes of background sources, can be easily included in the inference.
Moreover, we perform different functional tests, comparing a hierarchical approach with state-of-the-art zero-scatter models, to assess which sources of systematics can be resolved by this method, and show that the hierarchical models do not over-fit.

This paper is structured as follows. Our hierarchical inference is detailed in \Sec\ref{sec:hierarchical_cluster_lensing}. \Sec\ref{sec:technicalities} covers some technicalities inherent to our code implementation of the models. In \Sec\ref{sec:functional_tests}, we perform functional tests
on three simple but realistic mock clusters. Results are discussed in \Sec\ref{sec:results}, and we conclude in \Sec\ref{sec:conclusions}.

While a cosmological model is adopted to generate mock clusters and fit them, the main results of this work are general and separate from the choice of cosmology. In fact, if needed, cosmological parameters may be sampled as additional hyperparameters in our inference scheme.
   

\section{A Hierarchical cluster lensing model}
\label{sec:hierarchical_cluster_lensing}
In this section, we generalize the formalism of parametric cluster lensing models to allow for hierarchical inference, in particular scaling relations with non-zero scatter. Given the abundance of symbols, in the following (\Sec\ref{sec:measured_kinematics} and \ref{sec:posterior}) all model parameters and hyperparameters are marked by a ``hat'' symbol.

\subsection{Cluster lensing components}
In a parametric cluster lensing model, the total mass distribution is typically divided into a few component classes \citep{Jullo_lenstool}. Here, we indicate each class through its contribution, $\phi^{class}$, to the lensing potential. A first, smooth, cluster-scale component ($\phi^{halo}$) accounts for both the DM content of the cluster and the baryonic intra-cluster gas and light contributions. A second ``clumpy'' component describes the mass in cluster member halos ($\phi^{gal}$), in DM and baryons. A third component accounts for  the  presence  of  massive  structures  in  the  outer  cluster regions, and possibly additional massive halos along the line-of-sight ($\phi^{shear+los}$). Within this model class, the total cluster potential is then:

\begin{equation}
    \label{eq.: mass_decomposition}
    \phi_{tot} = \sum_{i=1}^{N^{h}} \phi_i^{halo}+ \sum_{j=1}^{N^{gal}_{tot}} \phi_j^{gal}+ \sum_{k=1}^{N^{sl}} \phi^{shear+los}_k,
\end{equation}

\noindent with $N^h$, $N^{gal}_{tot}$, $N^{sl}$, the number of cluster scale halos, cluster members, and shear plus line-of-sight contributions respectively.

The number of constraints given by the observed multiple image positions is usually not sufficient to fit the parameters of mass profiles used to describe each individual galaxy in a given cluster lens model. For this reason, cluster members are usually parameterized as circular dual pseudo-isothermal mass distributions (circular dPIEs, \citealt{Limousin_lenstool}, \citealt{Eliasdottir_lenstool}), with negligible core radii, whose parameters are related to galaxy luminosities according to fixed scaling relations. 

The circular dPIE is defined by the 3D mass density \citep{Limousin_lenstool}:

\begin{gather}
    \label{eq.: Density_3d_dPIE}
    \rho(r) =  \frac{\rho_0}{(1+r^2/r^2_{core})(1+r^2/r^2_{cut})},
\end{gather}

\noindent with:

\begin{gather}
    \label{eq.: Central_density_dPIE}
    \rho_0 =  \frac{\sigma^2_0}{2 \pi G} \frac{r_{cut}+r_{core}}{r_{core}^2 r_{cut}}.
\end{gather}

\noindent In these equations, $\sigma_{0}$, $r_{core}$, and $r_{core}$ are the central velocity dispersion, the core radius, and the truncation radius of the dPIE, respectively. For a vanishing $r_{core}$, $r_{cut}$ encircles about half of the total 3D dPIE mass and 60\% of the projected mass. In the limit $r_{cut}\rightarrow{\infty}$ and $r_{core}>0$, the dPIE coincides with the pseudo isothermal mass distribution defined in \cite{Kassiola_1993} (PIEMD).
In lensing models of galaxy clusters, the following scaling relations are commonly adopted for the central velocity dispersions ($\sigma_0^{gal}$), core radii ($r_{core}^{gal}$), and truncation radii ($r_{cut}^{gal}$) of cluster members:

\begin{equation}
\label{eq.: Scaling_relation_sigma}
    \sigma^{gal}_{0,i}= \sigma^{ref}_{0} \left(  \frac{L_i}{L_0} \right)^{\alpha}\ ,\ r^{gal}_{core,i}= r^{ref}_{core} \left(  \frac{L_i}{L_0} \right)^{\beta_{core}}\ ,\ r^{gal}_{cut,i}= r^{ref}_{cut} \left(  \frac{L_i}{L_0} \right)^{\beta_{cut}}\ ,
\end{equation}

\noindent without any intrinsic scatter.
Under the hypothesis of a power-law scaling $M_{tot,i}/{L_i}\propto L_i^\gamma$ of the total mass of a cluster member, and since the total dPIE mass is $M_{tot} = {(\pi \sigma_{0}^2 r_{cut})}/{G}$, 
the exponents are therefore related through:

\begin{equation}
    \beta_{cut}=\gamma-2\alpha+1\ .
    \label{eq.: slopes}
\end{equation}

\subsection{Measured kinematics}
\label{sec:measured_kinematics}

Spectroscopic information, when incorporated inside the lensing models, produces robust and accurate reconstructions of the projected cluster masses. 
However, strong lensing models still suffer from internal degeneracies between the parameters of their mass components (\Eq\ref{eq.: mass_decomposition}). A degeneracy exists, for example, between the two normalizations, $\sigma_0^{ref}$ and $r_{cut}^{ref}$ in \Eq\ref{eq.: Scaling_relation_sigma}.
This degeneracy is easily understandable if we consider that the total mass of a dPIE, with negligible $r_{core}$, inside an aperture of radius $R$ is \citep{Jullo_lenstool}:

\begin{equation}
\label{mass_deg}
    M(R)=\frac {\pi \sigma_0^2}{G}\left(R+r_{cut}-\sqrt{r_{cut}^2+R^2}\right),
\end{equation}

\noindent and that multiple images constrain only the total mass within their (projected) distance from the center of a galaxy. Other degeneracies also exist between the cluster-scale DM halo parameters and the clumpy subhalo components since the overall mass may be redistributed differently between the main halo and the subhalos, except where the multiple-image positions are strongly constraining.

Recently, velocity dispersions of cluster members (from fits to spectra with signal-to-noise ratio grater than ten) were used to break, or at least reduce, these internal degeneracies (\citealt{Verdugo_2007,Monna_2015,Monna_2017b} and more extensively by \citealt{ber19}). \cite{Monna_2015} and \cite{Monna_2017b} fixed the velocity dispersions of several cluster galaxies to their measured values, thereby breaking their model degeneracies by imposing a strong hypothesis on their mass content. In a more flexible approach, \citet{ber19} used high-quality MUSE spectra of cluster member galaxies to build priors on the parameters of the scaling relations in \Eq\ref{eq.: Scaling_relation_sigma}. 

In this work, we combine both of the above to fully characterize the subhalo components of clusters. Our purpose-built \textsc{BayesLens} wrapper uses the available measured velocity dispersions of the cluster galaxies, $\sigma_m^{gal}\pm\delta\sigma_m^{gal}$, to infer the hyperparameters ($\hat{\sigma}^{ref}$, $\hat{\alpha}$) of the $\sigma\mbox{-}mag$ scaling relation. 
A third hyperparameter, $\hat{\Delta\sigma}^{ref}$, quantifies the scatter of the measured galaxies around the $\sigma\mbox{-}mag$ scaling relation. Gaussian priors, centered on the measured $\sigma_m^{gal}$ and with standard deviations equal to five times the measured errors $\delta\sigma_m^{gal}$, are adopted in the lens model for the dPIE velocity dispersions of galaxies with measured velocity dispersions. For galaxies without a measured velocity dispersion, we assumed Gaussian priors centered on the inferred $\sigma\mbox{-}mag$ scaling relation and with a standard deviation equal to $\hat{\Delta\sigma}^{ref}$. We adopt a uniform prior for the reference cut radius of the $r_{cut}\mbox{-}mag$ scaling relation, and $\beta_{cut}$ is obtained from \Eq\ref{eq.: slopes}. Unless otherwise stated, all model hyperparameters are left free to vary to fully explore the model posterior probability as described below. All ``hat'' symbols are introduced more formally in the following subsection (\Sec\ref{sec:posterior}).

\subsection{Fitting it all together: the posterior}
\label{sec:posterior}

In our models, we use the measured velocity dispersions of $N_{m}^{gal}$ cluster galaxies, $\sigma_m^{gal}\pm\delta \sigma_m^{gal}$, together with the positions $\mathbf{x}_{im}$ of $N^{im}$ multiple images, from $N^{fam}$ different sources with positions $\mathbf{x}_{sou}$, as observational constraints to the lens model free-parameters. Hereafter, these free-parameters will be marked with a hat symbol. 
In order to explore the lens models, we sample the total posterior probability function $p_{tot}$, expressed as the product of: the odds $p_{g}$ of drawing each galaxy (independently) from a given scaling relation (with intrinsic scatter); the likelihood $p_{sr}$ of the scaling relation hyperparameters, given the measured velocity dispersions and luminosities; the likelihood $p_{im}$ of reproducing the observed image positions, given the cluster model (including the parameters of each cluster member galaxy); a prior $p_{h}$ on the parameters of the main cluster-scale halo(s). We also include a term $p_{mg}$ that links the dPIE lensing velocity dispersions of the galaxies with measured kinematics to their stellar velocity dispersions, with large uncertainties. Although not formally needed, this term is used only as a loose regularization to ensure convergence.

In synthesis,
\begin{multline}
    \label{eq.: posterior_tot}
    p_{tot}\left(\hat{\sigma}^{ref}, \hat{\alpha}, \hat{\Delta \sigma}^{ref},\hat{r}_{cut}^{ref},\hat{\sigma}_m^{gal},\hat{\sigma}^{gal},\hat{\phi}^{h} \mid mag^{gal}\sigma_m^{gal}, \delta \sigma_m^{gal},\mathbf{x}_{im}\right)\propto\\ \propto p_{sr}\left(\hat{\sigma}^{ref}, \hat{\alpha}, \hat{\Delta \sigma}^{ref}, \hat{r}_{cut}^{ref} \mid mag^{gal}\sigma_m^{gal}, \delta \sigma_m^{gal}\right) \times\\\times p_{mg}\left(\hat{\sigma}_{m}^{gal} \mid \sigma_m^{gal}, \delta \sigma_m^{gal}\right)\times p_{g}\left(\hat{\sigma}^{gal} \mid \hat{\sigma}^{ref}, \hat{\alpha}, \hat{\Delta \sigma}^{ref}\right)\times\\ \times p_{h}\left(\hat{\phi}^h\right)\times p_{im}\left( \mathbf{x}_{im}\mid \hat{\sigma}^{ref}, \hat{\alpha}, \hat{\Delta \sigma}^{ref},\hat{r}_{cut}^{ref},\hat{\sigma}_m^{gal},\hat{\sigma}^{gal},\hat{\phi}^{h},\mathbf{x}_{sou} \right).
\end{multline}
Each of the five factors on the right-hand-side is discussed below. In the following, the quantities referring to cluster galaxies with measured velocity dispersions are marked with the subscript ``m''.

Some of the factors (e.g., the one on hyperparameters) can be interpreted as a posterior on some parameters given some observations, which (by the ``Bayes chain rule'') can further be used as a prior for the full hierarchical posterior. To avoid confusion between those posteriors and the final posterior, we alternatively refer to them as ``term'' or ``odds'' in the following.

\subsubsection{Odds on the scaling relation hyperparameters, $p_{sr}$}

\noindent This factor is responsible for the $\sigma\mbox{-}mag$ and $r_{cut}\mbox{-}mag$ scaling relation hyperparameters, given the set of $N_{m}^{gal}$ measured cluster galaxies. For the galaxy velocity dispersions, we consider here a scaling relation of the same form of \Eq\ref{eq.: Scaling_relation_sigma} parameterized by the reference measured velocity $\hat{\sigma}^{ref}$ and slope $\hat{\alpha}$ plus an intrinsic scatter $\hat{\Delta \sigma}^{ref}$ in measured velocity dispersions. Regarding the $r_{cut}\mbox{-}mag$ scaling relation, in the current version of our models we optimize only the reference value $\hat{r}_{cut}^{ref}$ while the slope $\beta_{cut}$ is determined using \Eq\ref{eq.: slopes} from the inferred $\hat{\alpha}$ and assuming a fixed mass-to-light scaling for the cluster galaxies. No scatter around this relation is considered.

The term $p_{sr}$ can be expressed as:

\begin{multline}
    \label{eq.: posterior_scaling_relations}
    p_{sr} \left(\hat{\sigma}^{ref}, \hat{\alpha}, \hat{\Delta \sigma}^{ref},\hat{r}_{cut}^{ref} \mid mag^{gal}, \sigma^{gal}_{m}, \delta\sigma^{gal}_{m}\right) \propto \\ p_{0,sr} \left(\sigma^{gal}_{m} \mid mag^{gal}, \delta\sigma^{gal}_{m}, \hat{\sigma}^{ref}, \hat{\alpha}, \hat{\Delta \sigma}^{ref}\right) \times \\ \times \varpi_{sr} \left(\hat{\sigma}^{ref}, \hat{\alpha}, \hat{\Delta \sigma}^{ref},\hat{r}_{cut}^{ref} \right),
\end{multline}
\noindent where the uninformative prior $\varpi_{sr}$ is defined by:
\begin{equation}
    \label{eq.: prior_scaling_relations}
    \ln\left\{ \varpi_{sr}\left(\hat{\sigma}^{ref}, \hat{\alpha}, \hat{\Delta \sigma}^{ref}\right)\right\}=
    \begin{cases}
        - \ln(\hat{\Delta \sigma}^{ref}),& \scriptstyle \text{if } \sigma^{ref}_{min}<\hat{\sigma}^{ref}<\sigma^{ref}_{max}\\ & \scriptstyle \text{and } \alpha_{min}<\hat{\alpha}<\alpha_{max} \\ & \scriptstyle \text{and } \Delta \sigma^{ref}_{min}<\hat{\Delta \sigma}_{m}<\Delta \sigma^{ref}_{max}\\ &
        \scriptstyle \text{and }
        r^{ref}_{cut,min}<\hat{r}_{cut}^{ref}<r^{ref}_{cut,max}\\
        -\infty,              & \scriptstyle \text{otherwise}
    \end{cases},
\end{equation}

\noindent and limits the $\sigma\mbox{-}mag$ plus scatter scaling relation hyperparameters and the reference truncation radius, $\hat{r}_{cut}^{ref}$, to lie within suitably chosen boundaries $\sigma^{ref}_{min(max)}$, $\alpha_{min(max)}$, $\Delta\sigma^{ref}_{min (max)}$, and $r_{cut,min(max)}^{ref}$.
The log-likelihood term

\begin{multline}
    \label{eq.: likelihood_scaling_relations}
    \ln \left\{p_{0,sr}\left(\sigma_{m}^{gal} \mid mag^{gal}, \delta\sigma_{m}^{gal}, \hat{\sigma}^{ref}, \hat{\alpha}, \hat{\Delta \sigma}^{ref}\right)\right\} =\\= - \frac{1}{2} \sum_{i=1}^{N_{m}^{gal}} \left[ \frac{\left(\sigma_{m,i}^{gal}-\hat{\sigma}_{m,i}^{sr}\right)^2}{\left(\delta\sigma^{gal}_{m,i}\right)^2 + \left(\hat{\Delta \sigma}^{ref}\right)^2} + \right.\\ \left. +\ln \left\{2 \pi \left[\left(\delta\sigma^{gal}_{m,i}\right)^2 + \left(\hat{\Delta \sigma}^{ref}\right)^2 \right] \right\} \right],
\end{multline}

\noindent {quantifies the goodness-of-fit of scaling relation hyperparameters to the measured velocity dispersions.} In \Eq\ref{eq.: likelihood_scaling_relations}, we define $\hat{\sigma}_{m,i}^{sr}$ as:

\begin{equation}{
\hat{\sigma}_{m,i}^{sr}=\hat{\sigma}^{ref} 10^{\,0.4\,\hat{\alpha}\, \left(mag^{ref}-mag_i^{gal}\right)},
}
\label{eq.: scaling_model}
\end{equation}
\noindent where $mag^{ref}$ corresponds to the reference luminosity $L_0$ in \Eq\ref{eq.: Scaling_relation_sigma}.

\subsubsection{Term on measured galaxies, $p_{mg}$}
\noindent This term applies only to the $N_m^{gal}$ galaxies with measured velocity dispersions. It allows for residual uncertainties (e.g., from mass-anisotropy degeneracy and asphericity) in converting measured velocity dispersions in dPIE $\hat{\sigma}_{m}^{gal}$, which in turn propagate on the scaling relations that all galaxies are drawn from. This is attained by linking the kinematic sigma to the lensing sigma with some tolerance. If this were not done, one might obtain biased results if the kinematic sigma-mag relation has some intrinsic scatter, but the lensing sigma-mag does not. This term is also needed in order to account for systematic uncertainties in the ``measured'' dispersions, which in the high signal-to-noise ratio regime can dominate over the statistical uncertainties. Erring on the conservative side, we choose the conversion tolerance to five times the uncertainties in the measured velocity dispersion.

Therefore, $p_{mg}$ alone consists in Gaussian priors centering the cluster member aperture-average velocity dispersions $\hat{\sigma}_m^{gal}$ on their kinematic values $\sigma_m^{gal}$. We choose the standard deviation of Gaussian priors equal to five times the error
on the kinematic measurements $\delta \sigma_m^{gal}.$ This term is given by:
\begin{multline}
    \label{eq.: prior_measured_galaxies}
    \ln\left\{ p_{mg}\left(\hat{\sigma}_{m}^{gal} \mid \sigma_{m}^{gal}, 5\,\delta\sigma_{m}^{gal}\right)\right\}=\\= -\frac{1}{2} \sum_{i=0}^{N_{m}^{gal}}\left\{\frac{\left(\hat{\sigma}_{m,i}^{gal}-\sigma_{m,i}^{gal}\right)^2}{\left(5\delta\sigma_{m,i}^{gal}\right)^2}+\ln\left[ 2\pi\left(5\,\delta\sigma_{m,i}^{gal}\right)^2 \right]\right\}.
\end{multline}

In other words, this term attributes to galaxies their measured velocity dispersions, with very wide tolerance, unless a deviation from these values produces a significant improvement of the lensing model. We emphasize that this term is, strictly speaking, not necessary, and formally it may prevent the finite-scatter models from reducing to zero-scatter models (if this is needed by the data). However, given the typical $\delta\sigma_{m}^{gal}\gtrsim15$~km/s uncertainties on measured kinematics, the $5\,\delta\sigma_{m}^{gal}$ term ensures that this factor only acts as a loose regularization, preventing pathological solutions and aiding the convergence of the models.

\subsubsection{Prior on unmeasured galaxies, $p_g$}
\noindent This term is a collection of Gaussian priors on velocity dispersion values for the $N^{gal}=N^{gal}_{tot}-N_{m}^{gal}$ galaxies without kinematics measurements. Its form is such that the final posterior prefers lens models in which the unmeasured galaxies lie on the $\sigma\mbox{-}mag$ scaling relation inferred by $p_{sr}$, unless otherwise required by the lensing data:
\begin{multline}
    \label{eq.: prior_unm}
    \ln \left\{p_g\left(\hat{\sigma}^{gal} \mid \hat{\sigma}^{ref}, \hat{\alpha}, \hat{\Delta \sigma}^{ref}\right)\right\} =\\= - \frac{1}{2} \sum_{i=1}^{N^{gal}} \left\{ \frac{\left(\hat{\sigma}_{i}^{gal}-\hat{\sigma}_{i}^{sr}\right)^2}{\left(\hat{\Delta \sigma}^{ref}\right)^2} + \ln \left[2 \pi \left(\hat{\Delta \sigma}^{ref}\right)^2 \right] \right\}.
\end{multline}
\noindent The $\hat{\sigma}^{sr}$ are computed through \Eq\ref{eq.: scaling_model}, but now considering galaxies without measured velocity dispersions.

\subsubsection{Prior on halo parameters, $p_h$}
This term consists of flat priors on the smooth cluster-scale halo parameters collectively indicated as $\hat{\phi}^h$:

\begin{equation}
    \label{eq.: prior_halo}
    \ln\left\{ p_{h}\left(\hat{\phi}^{h}\right)\right\}=
    \begin{cases}
        0,& \text{if } \phi_{min}^{h}<\hat{\phi}^{h}<\phi_{max}^{h}\\
        -\infty,              &  \text{otherwise}
    \end{cases}
\end{equation}

\noindent If (as follows) these halos are parameterized as PIEMD profiles, $p_h$ is a prior on the sky coordinates $\hat{x}^{h}$ and $\hat{y}^{h}$, on the ellipticity $\hat{e}^{h}$, position angle $\hat{\theta}^h$, core radius $\hat{r}_{core}^h$, and central velocity dispersion $\hat{\sigma}_0^{h}$.

\subsubsection{Multiple-image likelihood, $p_{im}$}
\noindent This final term is a likelihood function quantifying the agreement between observed and lens model predicted multiple-image positions. 
Given $N^{fam}$ sources with $N^{im}_i$ multiple images associated to the same $i$-th source (usually called a ``family''), we follow \cite{Jullo_lenstool} and express the likelihood as:

\begin{multline}
    \label{eq.: pi}
p_{im}\left( \mathbf{x}_{im}\mid \hat{\sigma}^{ref}, \hat{\alpha}, \hat{\Delta \sigma}^{ref},\hat{r}_{cut}^{ref},\hat{\sigma}_m^{gal},\hat{\sigma}^{gal},\hat{\phi}^h,\mathbf{x}_{sou}\right) =\\= \prod^{N^{fam}}_{i=0} \frac{\mathrm{e}^{-\chi^2_i/2}}{\prod_j \Delta x_{i,j}\sqrt{2\pi}},
\end{multline}

\noindent where the $\chi^2_i$ associated to the $i$-th family is:

\begin{equation}
    \label{eq.: chisq_lenstool}
    \chi^2_i = \sum_{j=1}^{N^{im}_i} \frac{\left\| \mathbf{x}_{i,j}^{obs} - \mathbf{x}_{i,j}^{pred} \right\|^{2}}{\Delta x_{i,j}^2},
\end{equation}

\noindent with $\mathbf{x}^{obs}_{i,j}$ the observed positions of the multiple images on the lens plane, $\Delta x_{i,j}$ are the isotropic uncertainties on these positions and $\mathbf{x}_{i,j}^{pred}$ are the model predicted positions, given the adopted cosmology and the inferred set of model parameters: $\hat{\sigma}^{ref}, \hat{\alpha}, \hat{\Delta \sigma}^{ref},\hat{r}_{cut}^{ref},\hat{\sigma}_m^{gal},\hat{\sigma}^{gal},\hat{\phi}^{h}$.

\section{Technicalities}
\label{sec:technicalities}

To sample the complete posterior in \Eq\ref{eq.: posterior_tot}, we use the Affine-Invariant sampling as originally introduced by \citet{gw2010}, which is especially suited to our highly-dimensional (and possibly degenerate) parameter space. In particular, to enable full portability and reproducibility, we use the latest \texttt{python} release\footnote{https://github.com/dfm/emcee} of \texttt{emcee} \citep{emcee_2013}.

The first four terms in equation \Eq\ref{eq.: posterior_tot} are directly implemented in our code, while to compute the multiple-images likelihood $p_{im}$ we exploit the publicly available software \textsc{LensTool} (\citealt{Kneib_lenstool}, \citealt{Jullo_lenstool}, \citealt{Jullo_Kneib_lenstool}). 
The synergy between our code and \textsc{LensTool} requires some technicalities, as described in the following subsections. In any case, our code is fully modular, so that \textsc{LensTool} can be replaced by any other parametric lensing software by changing only a few \texttt{python} lines.

\subsection{Calls to \textsc{LensTool}: input and output files}
\label{sec:input_outputs}
For each parameters combination, corresponding to a given walker position inside the parameters space, \textsc{BayesLens} silently calls \textsc{LensTool} to compute the $p_{im}$ term of the total posterior. Every \textsc{LensTool} call needs a different input file generated in our code by a specific \texttt{python}-function. Another function reads the resulting likelihood, computed by \textsc{LensTool}, from an output file. Since all these output files are saved on the disk using the same name, we create folders with unique (random) names to differentiate each \textsc{LensTool} call. These folders are deleted at the end of every $p_{im}$ computation.

To sample the posterior $p_{tot}$, millions of walker positions are required. Thus, millions of input and output \textsc{LensTool} files are quickly created and deleted on the disk during this process. To avoid the disk wear and the bottleneck represented by the process of writing/reading files, part of our computer RAM is reserved for creating a ``RAM-disk'' where the input and output files are temporarily saved and deleted.

\begin{figure*}[ht!]
	\centering
	\includegraphics[width=\linewidth]{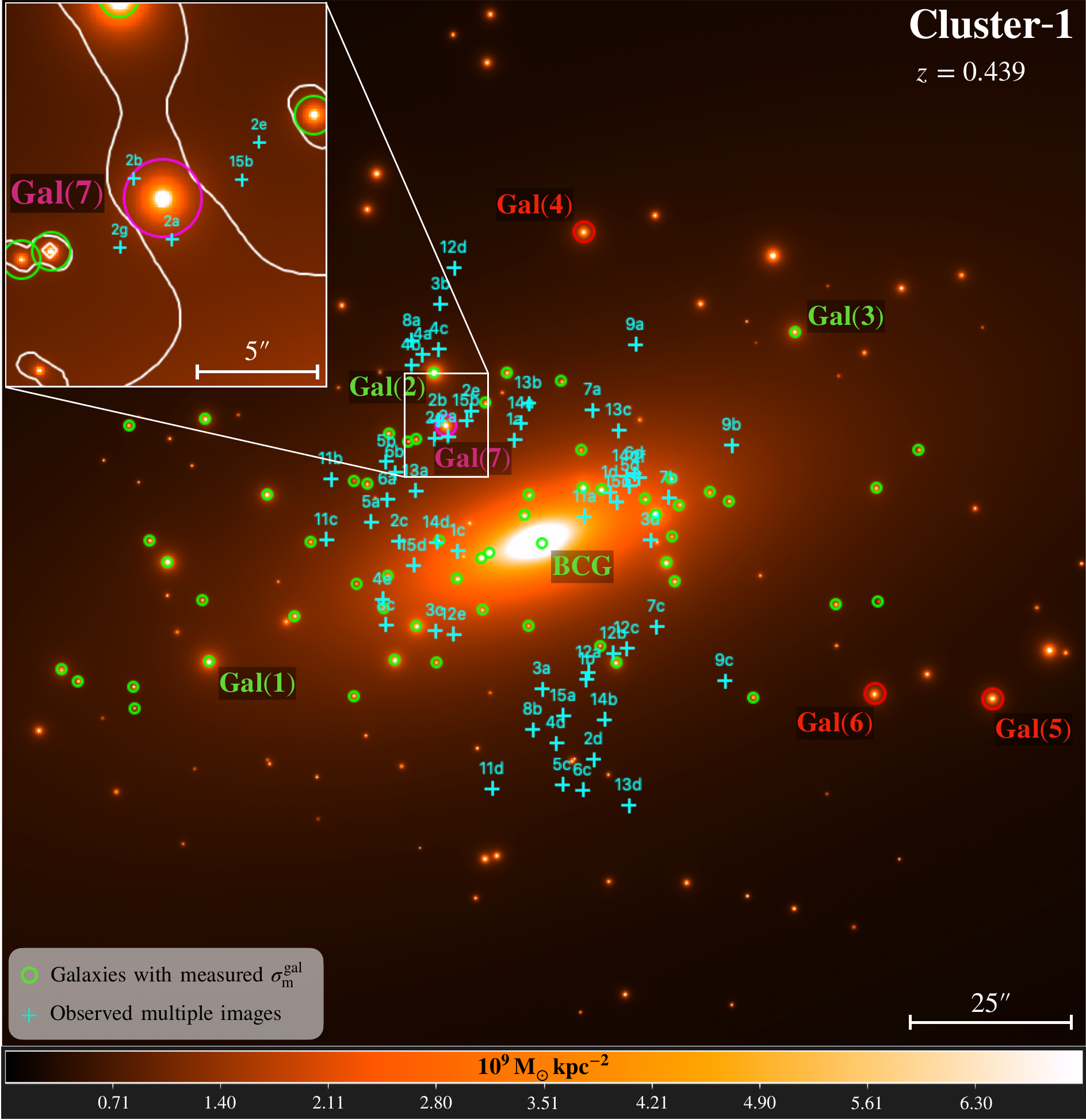}
	\caption{Mass density distribution, color coded in $\mathrm{M_{\odot}\,kpc^{-2}}$, of the central region of the Cluster-1 mock used for our functional tests, loosely based on the \cite{Caminha_macs1206} lens model for the cluster MACS J1206.2$-$0847 at redshift $z=0.439$. Green circles mark the galaxies for which we have a `measured' velocity dispersion. Our working hypothesis is that these velocity dispersions are measured within apertures of radius R=0.8\arcsec, displayed by the green circles. We label in green and red the brightest galaxies with and without 'measured' velocity dispersion, respectively. 
	Cyan crosses mark the position of 'observed' multiple images for which we assume an isotropic statistical error of 0.2\arcsec. In the inset, we show a galaxy-scale multiple-image system around the cluster member Gal(7) (labeled in magenta). The critical lines computed for the redshift, $z_2=2.539$, of the second family of multiple images are shown in white (see text).}
	\label{fig:mock_cluster}
\end{figure*}

\subsection{From measured $\sigma_{m}^{gal}$ to \textsc{LensTool} fiducial $\sigma_{LT}$}
\label{sec:velocity}
The dPIE is implemented in \textsc{LensTool} through a fiducial velocity dispersion $\sigma_{LT}$ related to the 1D central velocity dispersion $\sigma_0$ (in \Eq\ref{eq.: Central_density_dPIE}) by: $\sigma_0=\sqrt{3/2}\sigma_{LT}$. 
To convert the model predicted aperture-average line-of-sight velocity dispersions $\hat{\sigma}_m^{gal}$ and $\hat{\sigma}^{gal}$ to their fiducial \textsc{LensTool} values ($\hat{\sigma}_{(m),LT}^{gal}$), we relate them through

\begin{gather}
    \label{eq.: LensTool_sigma}
    \hat{\sigma}_{(m),LT}^{gal}=\hat{\sigma}_{(m)}^{gal}/c_p(R)\ ,
\end{gather}

\noindent where $R$ is the aperture radius chosen for the cluster member spectral extraction.
Adopting orbital isotropy and a (spherical) galaxy surface brightness profile proportional to the dPIE matter density, the projection coefficient $c_p(R)$ is given by:

\begin{multline}
    \label{eq.: projection_coefficient}
    c_p^2(R) = \frac{6}{\pi}\frac{r_{c}+r_{t}}{r_{c}^2 r_{t}}
     \left( \sqrt{r_{c}^2+R^2}-r_{c}-\sqrt{r_{t}^2+R^2}+r_{t}\right)^{-1}\times \\
    \times \int_{0}^{R}\int_{R'}^{\infty} R'\frac{ r_{t} \arctan \left(\frac{r}{r_{t}}\right)-r_{c} \arctan \left(\frac{r}{r_{c}}\right)}{(1+r^2/r^2_{c})(1+r^2/r^2_{t})}\frac{\sqrt{r^2-R'^2}}{r^2}drdR',
\end{multline}

\noindent with $r_{c}=r_{core}$ and $r_{t}=r_{cut}$ \citep[see,][]{ber19}.

\section{Functional tests}
\label{sec:functional_tests}

To test the ability of \textsc{BayesLens} in recovering the correct halos and subhalos mass parameters and in predicting the correct multiple-image positions, we develop three mock galaxy clusters, namely Cluster-1, Cluster-2, and Cluster-3, with an increasing degree of complexity. Cluster-1 is a simple and well-controlled cluster mock, while Cluster-2 and 3 have the main scope to analyze the role of low-mass cluster substructures and accuracy in the multiple-image positions. Although none of the models claim to reproduce the complexity of real galaxy clusters, they represent ideal preliminary tests for our code, leaving to future works its application to more complex simulations \citep[see][]{Meneghetti_2017} and to real cluster observations. In particular, the mocks allow us to verify the flexibility of our hierarchical approach, their dependence on the lensing or kinematics constraints, and their robustness against over-fitting. Since our simple mocks are developed using \textsc{LensTool}, the results we obtain are directly comparable to model inputs, thus avoiding the possibility of parameter misidentifications in the discussion of the results.

In this section, we describe the main characteristics of the simulated clusters and the settings adopted in the \textsc{BayesLens} optimization. We emphasize that these are functional tests, that is, on the behavior of the method itself in recovering the input parameters of simple but realistic mocks. These are the very first tests that any new method undergoes before it is tested on less-well controlled mocks (e.g., from simulations) to ensure that it can properly reproduce its own input parameters. On fully simulated or real-life clusters, the performance of each separate method may worsen because the true underlying mass distribution is unknown ``a priori'', so the only meaningful comparison is among different models on well-controlled mocks.

\subsection{Cluster-1: the simplest toy cluster model}
\label{sec:model1}

Cluster-1 is based on the best-fit \textsc{LensTool} lens model developed by \cite{Caminha_macs1206} (hereafter \ca) for the CLASH cluster MACS J1206.2$-$0847 at redshift $z=0.439$. With respect to \ca, the cluster-scale component of our simulated cluster is made by a single halo, parametrized through a PIEMD mass profile. The center of the halo has an offset of 0.92\arcsec\ from the brightest cluster galaxy (BCG) reference position. It has a core radius $r_{core}^h=3$\arcsec, while its \textsc{LensTool} fiducial velocity dispersion has a value of $\sigma_{LT}^h=1000$\vel. Its ellipticity $e=(a^2-b^2)/(a^2+b^2)$ is fixed to $e^h = 0.7$, with a position angle of $\theta^h = 19.14^{\circ}$ counterclockwise from west. 

\begin{table}[t!]     
	\tiny
	\def\arraystretch{1.6}
	\centering    
		\begin{tabular}{|c|c|c|c|c|}
	\cline{2-4}
		\multicolumn{1}{c|}{} & \multicolumn{3}{c|}{ {\bf{Cluster proprieties}}} \\
		\cline{2-4}
		  \multicolumn{1}{c|}{} & \bf{Cluster\mbox{-}1} & \bf{Cluster\mbox{-}2} &  \bf{Cluster\mbox{-}3} \cr 
		  
          \hhline{-===-}
          
		  {\boldmath{$N_m^{gal}$}} & 58 & 58 & 58 & \multirow{4}{*}{\rotatebox[origin=c]{270}{\textbf{SUBHALOS}}}  \cr \cline{1-4}
		  \boldmath{$N^{gal}(\mathrm{Out\, SR})$} & 80 (80) & 200 (22) & 200 (22) & \cr \cline{1-4} 
		  \boldmath{$N^{DH}$} & 0 & 0 & 910 &  \cr \cline{1-4}
		  \boldmath{$(\Delta x^{gal})_{sys}\,[\arcsec]$} & 0.0 & 0.01 & 0.01 & \cr \cline{1-4}
		  
		  \hhline{=====}
		  
		  \boldmath{$N^{im}_{tot}$} & 58 & 69 & 69 & \multirow{3}{*}{\rotatebox[origin=c]{270}{\textbf{IMAGES}}} \cr \cline{1-4}
		  \boldmath{$(\Delta x^{im})_{st}\,[\arcsec]$} & 0.2 & 0.2 & 0.2 & \cr \cline{1-4}
		  \boldmath{$(\Delta x^{im})_{sys}\,[\arcsec]$} & 0.0 & 0.01 & 0.01 & \cr \cline{1-4}
		  
		  \hhline{=====}
		  
		  \boldmath{$\chi^{2}_{tot}$} & 8.66 & 12.53 &  13.75 & \multirow{2}{*}{\rotatebox[origin=c]{270}{\textbf{RES.}}} \cr \cline{1-4}
		  \boldmath{$\Delta_{rms}^{tot}\,[\arcsec]$} & 0.08 & 0.09 & 0.09 & \cr \cline{1-5}
		  
	\end{tabular}
	
	\vspace{5mm}
	
    \caption{{Main features of the Cluster-1, Cluster-2, and Cluster-3 mocks.
    The number of galaxies with and without a measured velocity dispersion is given by $N_{m}^{gal}$ and $N^{gal}$ respectively. For the latter, we also quote (in brackets) the number of galaxies, in \textsc{BayesLens}, {that in principle are} free to scatter around the best-fit $\sigma\mbox{-}mag$ scaling relation. The systematic uncertainty assumed on galaxy positions is $(\Delta x^{gal})_{sys}$. $N^{im}_{tot}$ is the total number of observed multiple images, while $(\Delta x^{im})_{st}$ and $(\Delta x^{im})_{sys}$ are the statistical and systematic isotropic errors assumed on their positions. In the last two lines, we quote the total chi-square and the r.m.s. displacement between the ``measured'' and model-predicted image positions.
    }}    

	\label{table:model_ch}

\end{table}

\begin{figure*}[h!]
	\centering
	\includegraphics[width=\linewidth]{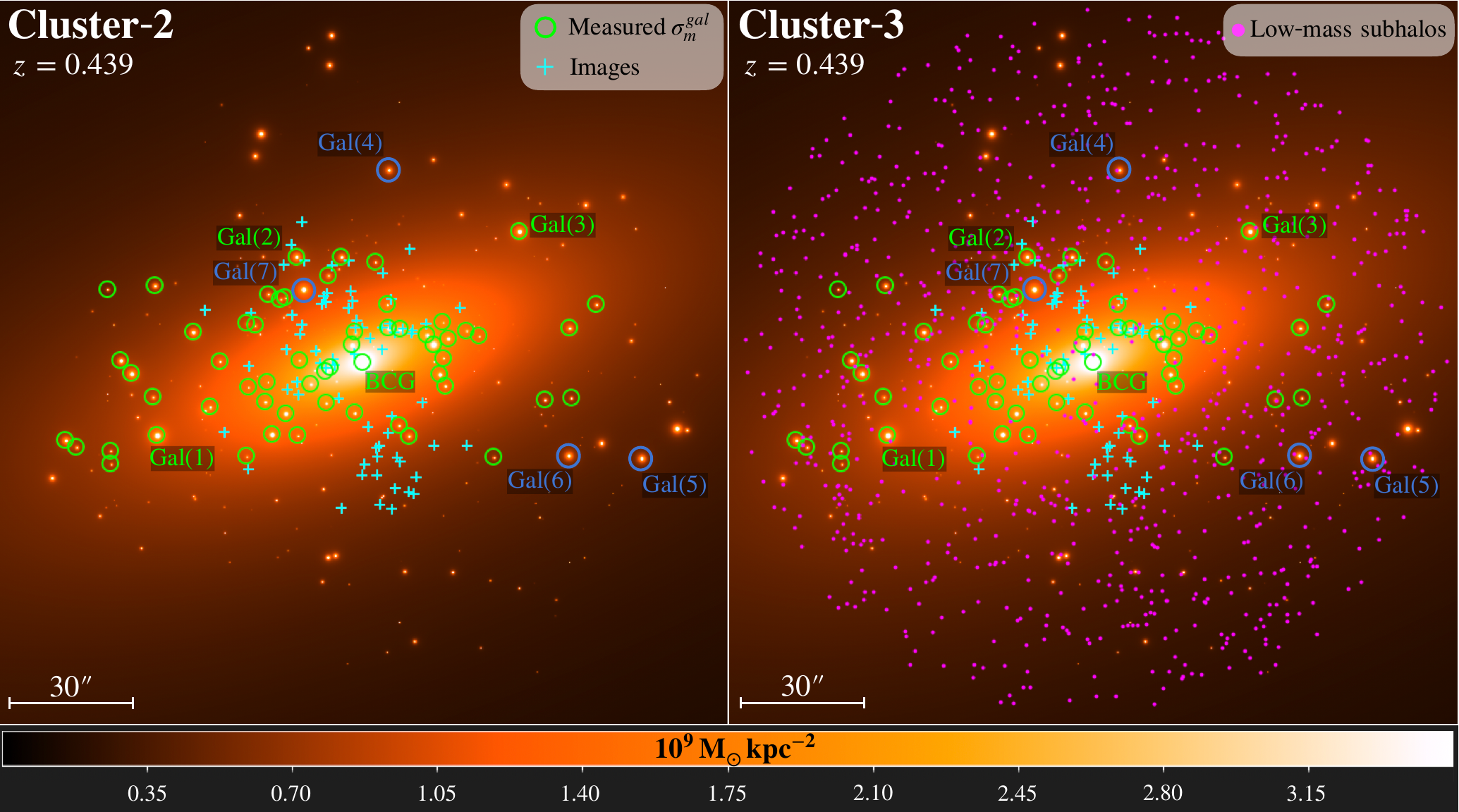}
	\caption{Mass density distribution, color-coded in $\mathrm{M_{\odot}\,kpc^{-2}}$, of the central region of the Cluster-2, and Cluster-3 mocks. As in Fig.\,\ref{fig:mock_cluster}, green circles mark the galaxies for which we have a ``measured'' velocity dispersion, while cyan crosses are the ``observed'' multiple images. We label in green and blue the same set of measured and unmeasured galaxies selected in Fig.\,\ref{fig:mock_cluster}. On the right panel, we plot magenta data-points showing the spatial distribution of the low-mass subhalo population of Cluster-3.}
	\label{fig:DH}
\end{figure*}

\begin{figure*}[h!] 
	\centering
	\includegraphics[width=\textwidth]{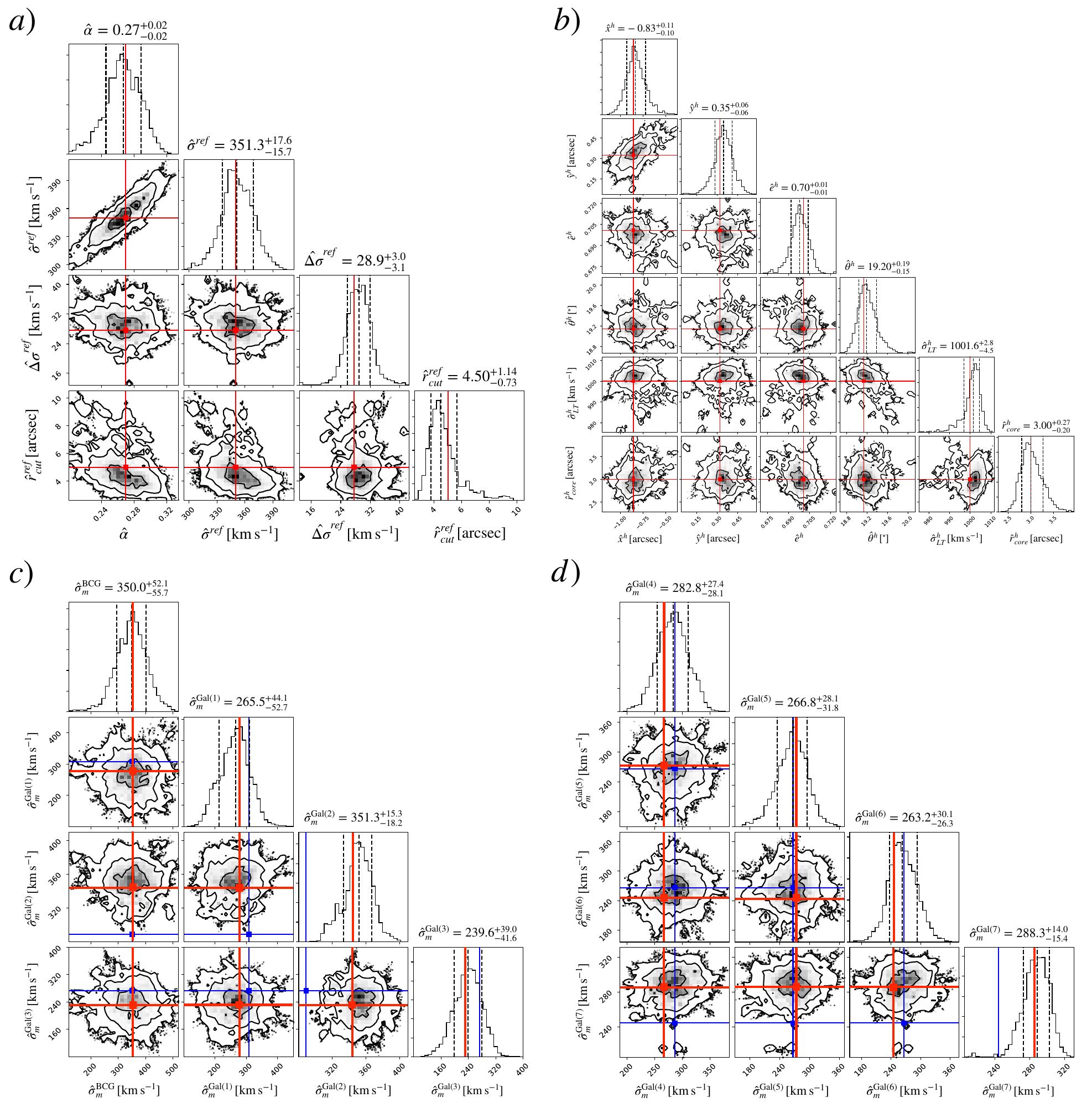}
	\caption{\textsc{BayesLens} marginalized posterior distributions on the free-parameters of the Cluster-1 mocks. The contours limit the $1\sigma$, $2\sigma$ and $3\sigma$ regions. The black dashed vertical lines in the histograms correspond to the 16th, 50th, and 84th percentiles of the marginalized distributions (these values are reported in the titles). The solid red lines are the 'true' values of the parameters of the mock cluster. In panel $\boldsymbol{a}$, the scaling relation hyperparameters are shown. In this plot we omit $\hat{\beta}_{cut}$ because its value is not optimized in \textsc{BayesLens} but directly derived from \Eq\ref{eq.: slopes} assuming $\gamma=2$. Panel $\boldsymbol{b}$ shows the posterior distributions for the cluster-scale halo parameters, while in panel $\boldsymbol{c}$ we show the average velocity dispersions within apertures of radius $R=0.8$\arcsec\ of the four brightest 'measured' galaxies: BCG, Gal(1), Gal(2) and Gal(3). Finally, panel $\mathbf{d}$ refers to the average velocity dispersions within apertures of radius $R=0.8$\arcsec\ of the three brightest galaxies without a 'measured' $\sigma$: Gal(4), Gal(5), and Gal(6). In the plot, we also include the galaxy Gal(7), forming a galaxy-scale strong lensing system (see text). In panels $\boldsymbol{c}$ and $\boldsymbol{d}$, we mark with blue vertical lines the stellar, aperture-averaged, velocity dispersion of cluster members predicted by the best-fit $\sigma^{gal}\mbox{-}m_{\mathrm{F160W}}$ scaling relation (zero-scatter solution).}
	\label{fig:degeneracies}
\end{figure*}

\begin{table*}[t!]     
	\tiny
	\def\arraystretch{1.6}
	\centering          
	\begin{tabular}{|l|c|c|c|c|}
	\cline{2-5}
		\multicolumn{1}{c|}{} & \multicolumn{4}{c|}{ {\bf{Scaling relation hyperparameters}} \boldmath{$(m_\mathrm{F160W}^{ref}=17.19)$}} \\[2pt]
		\cline{2-5}
		  \multicolumn{1}{c|}{} & \boldmath{$\hat{\alpha}$} & \boldmath{$\hat{\sigma^{ref}}\, \mathrm{[km\ s^{-1}]}$} & \boldmath{$\hat{\Delta \sigma}^{ref}\, \mathrm{[km\ s^{-1}]}$} & \boldmath{$\hat{r}^{ref}_{core}\, \mathrm{[arcsec]}$} \cr
          \hhline{-====}

		  {\bf True (Cluster-1)} & $0.27$ & $350.0$ & $27.6$ & $5.00$ \cr \cline{1-5}
		  {\bf BL  (Cluster-1)} & $0.27_{-0.02}^{+0.02}$ & $351.3_{-15.7}^{+17.6}$ &  $28.9_{-3.1}^{+3.0}$ &  $4.50_{-0.73}^{+1.14}$ \\[2pt]
		  \hhline{=====}
    
		  {\bf True  (Cluster-2,3)} & $0.28$ & $295.5$ & $22.7$ & $3.74$ \cr\cline{1-5}
		  {\bf BL (Cluster-2)} & $0.28_{-0.02}^{+0.02}$ & $289.9_{-11.2}^{+11.5}$ &  $19.3_{-2.1}^{+2.2}$ &  $3.91_{-0.61}^{+0.78}$ \cr\cline{1-5}
		  {\bf BL (Cluster-3)} & $0.28_{-0.02}^{+0.02}$ & $290.1_{-12.0}^{+10.8}$ &  $19.3_{-2.1}^{+2.6}$ &  $3.93_{-0.64}^{+0.77}$ \cr\cline{1-5}
	\end{tabular}
	
	\vspace{5mm}
	
	\begin{tabular}{|l|c|c|c|c|c|c|}
	\cline{2-7}
		\multicolumn{1}{c|}{} & \multicolumn{6}{c|}{ {\bf{Halo parameters}}} \\
		\cline{2-7}
		  \multicolumn{1}{c|}{} & \boldmath{$\hat{x}^h\, \mathrm{[arcsec]}$} & \boldmath{$\hat{y}^h\, \mathrm{[arcsec]}$} &  \boldmath{$\hat{e}^h$} & \boldmath{$\hat{\theta}^h\, \mathrm{[^{\circ}]}$} &  \boldmath{$\hat{\sigma}^h_{LT}\, \mathrm{[km\ s^{-1}]}$} & \boldmath{$\hat{r}^h_{core}\, \mathrm{[arcsec]}$} \\[2pt] 
          \hhline{-======}
		  {\bf True (Cluster-1)} & $-0.86$ & $0.32$ & $0.70$ & $19.14$ & $1000.0$ & $3.00$ \\[2pt] \cline{1-7}
		  {\bf BL (Cluster-1)} & $-0.83_{-0.10}^{+0.11}$ & $0.35_{-0.06}^{+0.06}$ & $0.70_{-0.01}^{+0.01}$ & $19.20_{-0.15}^{+0.19}$ & $1001.6_{-4.5}^{+2.8}$ & $3.00_{-0.20}^{+0.27}$ \\[2pt] \cline{1-7}
		  
		  \hhline{=======}
		  {\bf True (Cluster-2,3)} & $-1.40$ & $0.14$ & $0.72$ & $19.80$ & $1080.0$ & $11.00$ \\[2pt] \cline{1-7}
		  {\bf BL (Cluster-2)} & $-1.33_{-0.16}^{+0.16}$ & $0.16_{-0.07}^{+0.07}$ & $0.72_{-0.004}^{+0.004}$ & $19.77_{-0.11}^{+0.12}$ & $1079.0_{-3.0}^{+2.5}$ & $11.00_{-0.17}^{+0.18}$ \\[2pt] \cline{1-7}
		  {\bf BL (Cluster-3)} & $-1.34_{-0.16}^{+0.17}$ & $0.17_{-0.07}^{+0.08}$ & $0.72_{-0.005}^{+0.004}$ & $19.76_{-0.12}^{+0.11}$ & $1079.3_{-2.8}^{+2.7}$ & $11.02_{-0.17}^{+0.17}$ \\[2pt] \cline{1-7}
	\end{tabular}
	
	\vspace{5mm}
	
	\begin{tabular}{|l|c|c|c|c|}
	\cline{2-5}
		\multicolumn{1}{c|}{} & \multicolumn{4}{c|}{ \bf{Galaxies with measured \boldmath{$\sigma\ \mathrm{[km\ s^{-1}]}$}}} \\
		\cline{2-5}
		  \multicolumn{1}{c|}{} & \boldmath{$\hat{\sigma}^{\mathrm{BCG}}_{m}\ (m_{\mathrm{F160W}}=17.19)$} & \boldmath{$\hat{\sigma}^{\mathrm{Gal(1)}}_{m}\ (m_{\mathrm{F160W}}=17.72)$} &  \boldmath{$\hat{\sigma}^{\mathrm{Gal(2)}}_{m}\ (m_{\mathrm{F160W}}=17.98)$} & \boldmath{$\hat{\sigma}^{\mathrm{Gal(3)}}_{m}\ (m_{\mathrm{F160W}}=18.21)$} \cr 
          \hhline{-====}
		  {\bf True (Cluster-1)} & $354.0\pm 12.9$ & $278.2\pm 10.4$ & $344.1\pm 12.9$ & $230.8\pm 8.8$ \cr \cline{1-5}
		  {\bf BL (Cluster-1)} & $350.0_{-55.7}^{+52.1}$ & $265.5_{-52.7}^{+44.1}$ & $351.3_{-18.2}^{+15.3}$ & $239.6_{-41.6}^{+39.0}$ \\[2pt]
		  \hhline{=====}
		  {\bf True (Cluster-2,3)} & $292.9\pm 10.6$ & $261.2\pm 9.7$ & $202.0\pm 7.6$ & $233.8\pm 8.9$ \cr \cline{1-5}
		  {\bf BL (Cluster-2)} & $290.5_{-15.8}^{+15.6}$ & $252.3_{-42.8}^{+44.5}$ & $208.3_{-16.4}^{+16.8}$ & $233.1_{-43.1}^{+41.6}$ \cr \cline{1-5}
		  {\bf BL (Cluster-3)} & $292.1_{-16.4}^{+15.2}$ & $256.8_{-45.6}^{+46.2}$ & $209.4_{-17.1}^{+16.8}$ & $234.9_{-42.4}^{+41.2}$ \cr \cline{1-5}
	\end{tabular}
	
	\vspace{5mm}
	
	\begin{tabular}{|l|c|c|c|c|}
	\cline{2-5}
		\multicolumn{1}{c|}{} & \multicolumn{4}{c|}{ \bf{Galaxies without measured \boldmath{$\sigma\ \mathrm{[km\ s^{-1}]}$}}} \\
		\cline{2-5}
		  \multicolumn{1}{c|}{} & \boldmath{$\hat{\sigma}^{\mathrm{Gal(4)}}\ (m_{\mathrm{F160W}}=18.01)$} & \boldmath{$\hat{\sigma}^{\mathrm{Gal(5)}}\ (m_{\mathrm{F160W}}=18.30)$} &  \boldmath{$\hat{\sigma}^{\mathrm{Gal(6)}}\ (m_{\mathrm{F160W}}=18.31)$} &  \boldmath{$\hat{\sigma}^{\mathrm{Gal(7)}}\ (m_{\mathrm{F160W}}=18.66)$}\cr 
          \hhline{-====}
		  {\bf True (Cluster-1)} & $266.6$ & $274.1$ & $244.9$ & $285.5$ \cr \cline{1-5}
		  {\bf BL (Cluster-1)} & $282.8_{-28.1}^{+27.4}$ & $266.8_{-31.8}^{+28.1}$ & $263.2_{-26.3}^{+30.1}$ & $288.3_{-15.4}^{+14.0}$\\[2pt]
		  
		  \hhline{=====}
		  {\bf True (Cluster-2,3)} & $174.3$ & $217.6$ & $210.4$ & $256.8$ \cr \cline{1-5}
		  {\bf BL (Cluster-2)} & $231.6_{-20.6}^{+20.9}$ & $218.9_{-20.8}^{+20.3}$ & $218.4_{-19.7}^{+19.0}$ & $216.1_{-20.3}^{+20.4}$\cr \cline{1-5}
		  {\bf BL (Cluster-3)} & $233.3_{-20.5}^{+19.4}$ & $217.8_{-19.9}^{+20.2}$ & $217.1_{-19.9}^{+20.8}$ & $225.6_{-16.3}^{+16.6}$\cr \cline{1-5}
	\end{tabular}
	
	\vspace{5mm}
	
    \caption{Comparison between the ‘true’ input values of the {Cluster-1, Cluster-2, and Cluster-3 mocks {and} the \textsc{BayesLens} output results}. The 50th, 16th, and 84th percentiles computed from the marginalized posteriors are quoted. The first table refers to scaling relations hyperparameters, including the scatter of cluster galaxies around the inferred $\sigma\mbox{-}mag$ relation. In the second table, we show the cluster-scale halo parameters. The third table shows the results for the four brightest galaxies with a 'measured' velocity dispersion: BCG, Gal(1), Gal(2), and Gal(3). 
    Finally, the last table shows the results for the galaxy Gal(7) (see text) and for the three brightest galaxies without a 'measured' velocity dispersion: Gal(4), Gal(5), and Gal(6). The galaxy magnitudes are shown within round brackets.}    

	\label{table:input_output}

\end{table*}

The clumpy subhalo component of the mock cluster is composed by 138 cluster galaxies, selected from the \ca\ members catalog to have magnitudes in the HST/F160W filter given by $m_{\mathrm{F160W}}<22$. All these galaxies are parametrized as circular dPIE profiles whose {$r_{core}^{gal}$, $r_{cut}^{gal}$, and $\sigma_{LT}^{gal}$} values are determined as follows. For {$r_{core}^{gal}$ and $r_{cut}^{gal}$} we adopt the scaling relations in \Eq\ref{eq.: Scaling_relation_sigma} 
with slopes $\beta_{core}=0.5$ and $\beta_{cut}=0.66$. The two normalizations, computed at the BCG luminosity $L_0$ ($m_{\mathrm{F160W}}^{ref}=17.19$), are $r_{core}^{ref}=0.01\arcsec$ and $r_{cut}^{ref}=5\arcsec$. We assign line-of-sight stellar velocity dispersions to the cluster member galaxies, averaged within a circular aperture, assuming a 15\% Gaussian scatter around the scaling relation in \Eq\ref{eq.: Scaling_relation_sigma} with $\alpha=0.27$ and normalization equal to 350\vel\ at the BCG luminosity. To determine the \textsc{LensTool} fiducial velocity dispersions, we deproject the aperture-averaged velocity dispersions using \Eq\ref{eq.: LensTool_sigma} and assuming apertures of radius $R=0.8$\arcsec. No shear nor foreground structures are present in our Cluster-1 mock.

Given the total mass distribution for the mock cluster, we use \textsc{LensTool} to ray-trace the position of 15 background sources, randomly selected from the \ca\ catalog, to their multiple images on the lens plane. The sources are within a redshift range of $1.01\leq z\leq6.06$ and produce a total of
85 multiple images. Images are identified using an integer number and a letter in such a way that images with identical numbers in their ID belong to the same family. 26 of the 85 multiple images are excluded from the final sample, either because they are de-magnified or because they are too close to a cluster member to be detectable, in real clusters, due to light contamination. We also exclude one of the 15 families of multiple images (i.e., the 10th family adopting image IDs in Fig.\,\ref{fig:mock_cluster}) since it is constituted by a single lensed image. Therefore, our final mock multiple images catalog consists in 58 multiple images, from 14 background sources, shown in Fig.\,\ref{fig:mock_cluster}. The final simulated cluster, despite a purposely simple mass distribution, contains a number of halos, subhalos, and multiple images comparable to those of some CLASH or HFF clusters. 

In our tests, we consider isotropic statistical errors of 0.2\arcsec on the multiple-image positions. This is a compromise choice between the sharper accuracies that can be achieved (in principle) with state-of-the-art imaging and the tolerance that the models need in order to efficiently explore the full parameter space, avoiding narrow likelihood peaks around local minima. Choosing a $0.2\arcsec$ tolerance on the image positions also allows us to keep conservative uncertainties on the posteriors in the hyperparameters. In practice, one may apply these models in subsequent iterations where the tolerance is gradually reduced to match the astrometric uncertainties.

We also suppose that the stellar velocity dispersions of 58 luminous cluster galaxies are measured within apertures of radius $R=0.8\arcsec$.
This is comparable to the number of velocity dispersions that \cite{ber19} measured from the MUSE data-cube of MACS J1206.2$-$0847.
To associate an error to these simulated measures, we use the following empirical relation derived from the measurements by \citet{ber19}:

\begin{multline}
    \label{eq.: error}
    \frac{\delta\sigma_m^{gal}}{\sigma_m^{gal}} = 1.6\times 10^{-3}\,m_{\mathrm{F160W}}^3-8.53\times10^{-2}\,m_{\mathrm{F160W}}^2\\
    +1.509\,m_{\mathrm{F160W}}-8.879
\end{multline}

\subsection{Cluster-2 and Cluster-3: harder rungs with low-mass substructures and systematics in the astrometry of cluster galaxies and multiple images}
\label{sec:model23}

Similarly to Cluster-1, the cluster-scale component of both Cluster-2 and Cluster-3 mocks is described through a single elliptical PIEMD profile. The PIEMD has an off-set of about $1.41\arcsec$ from cluster BCG and it is characterized by an ellipticity $e^{h}=0.72$, a position angle of $\theta^{h}=19.8^{\circ}$, a core radius $r_{core}^{h}=11\arcsec$, and a \textsc{LensTool} fiducial velocity dispersion $\sigma_{LT}^{h}=1080\,\mathrm{km\,s^{-1}}$.
The galaxy-scale component of the new mocks extends over a wider magnitude range from the BCG luminosity ($m_{\mathrm{F160W}}^{ref}=17.19$) down to $m_{\mathrm{F160W}}=24$, and amounts to 258 galaxies extracted from the cluster member catalog by \ca. Cluster galaxies are modeled as circular dPIEs whose $r_{core}^{gal}$, $r_{cut}^{gal}$, and $\sigma_{LT}^{gal}$ are determined as in \Sec\ref{sec:model1} but assuming $\alpha=0.28$, $\sigma^{ref}=295.5\,\mathrm{km\,s^{-1}}$, $\beta_{cut}=0.64$, and $r_{cut}^{ref}=3.74\arcsec$. As in Cluster-1, a 15\% Gaussian scatter around the $\sigma^{gal}\mbox{-}m_{\mathrm{F160W}}^{gal}$ scaling relation is assumed for the line-of-sight projected, aperture average, stellar velocity dispersion of galaxies. With this choice of dPIE parameter values, the cluster member population in Cluster-2 and Cluster-3 ranges from a total mass of $1.6\times 10^{12}\,\mathrm{M_{\odot}}$, for the cluster BCG, down to a total mass of $1.1\times 10^{9}\,\mathrm{M_{\odot}}$, for the less massive galaxy of the cluster.
In the following, we assume to know the measured stellar velocity dispersion and associated errors of 58 cluster galaxies (see \Sec\ref{sec:model1}). These velocities are given as inputs to \textsc{BayesLens} for the lens model optimization.

What differentiates Cluster-2 from Cluster-3 is that in the latter model, we introduce, in addition to cluster galaxies, a large population of low mass subhalos inside the simulated cluster. Thus, Cluster-3 is mainly designed to test the impact of numerous, undetected, substructures on \textsc{BayesLens} performances. To assign a mass to the faint subhalos, we exploit the subhalo mass function fitted by \cite{Giocoli_2010a} and adopted by the MOKA algorithm for simulating the gravitational lensing by galaxy clusters \citep[see][]{Giocoli_2012}. In using this formula, we assume a virial mass of $M_{vir}=1.59 \times 10^{15}\,\mathrm{M_{\odot}}$ for the host cluster halo. This value corresponds to the total mass of MACS~J1206-0847 (\ca) contained within a distance from the cluster center of $r_{200c}=2.06\,\mathrm{Mpc}$ ($362.79\arcsec$ at $z=0.439$) that is expected to be close to the virial radius of the cluster \citep{Umetsu_2014}. Note that, a circularly symmetric PIEMD mass distribution with $r_{core}^{h}=11\arcsec$ and $\sigma_{LT}^{h}=1080\,\mathrm{km\,s^{-1}}$ has a mass, within a radius of 362.79\arcsec, exactly equal to $1.59 \times 10^{15}\,\mathrm{M_{\odot}}$.

Low mass subhalos are spatially distributed according to the function derived by \citet{Gao_2004} fitting the results of cosmological simulations. In particular, we adopt a concentration parameter $c_{200c}=r_{200c}/r_s=3.5$ for the host cluster halo and we assume $a=1.944$ in \Eq3 of \citet{Gao_2004}. Recent work by \cite{Meneghetti_2020} shows a detailed comparison between subhalos in state-of-the-art N-body and hydrodynamical simulations and observed subhalos from a sizable sample of cluster lens models.

In Cluster-3, we consider all the faint substructures within a distance of $89.65\arcsec$ from the cluster BCG (this is the distance of the most external galaxy considered in the model), and with masses ranging between $1.56 \times 10^{9}\,\mathrm{M_{\odot}}$, that is about the mass of the least massive cluster member, down to $0.93 \times 10^{9}\,\mathrm{M_{\odot}}$. Under these assumptions, the low mass subhalo population of the Cluster-3 mock amounts to 910 profiles. 

Each low-mass subhalo in the mock is parameterized as a circular dPIE, and to obtain a value for their $\sigma_{LT}$ and $r_{cut}$ parameters, we adopt the following procedure. Firstly, we compute the total mass for all the simulated cluster member galaxies ($M_{tot}^{gal}$). Then, we fit a $M_{tot}^{gal}\mbox{-}m^{gal}_{\mathrm{F160W}}$
relation between galaxy masses and magnitudes. This relation is adopted to associate a fictitious magnitude to each low-mass subhalo. Finally, we use scaling relations in \Eq\ref{eq.: Scaling_relation_sigma} to derive velocity dispersions, core radii, and truncation radii for the simulated subhalos from their inferred magnitudes. As for cluster galaxies, also for the low-mass subhalos, we assume 15\% Gaussian scatter around the $\sigma^{gal}\mbox{-}m_{\mathrm{F160W}}^{gal}$ scaling relation.

Using Cluster-2 and Cluster-3 mocks, we map the position of 22 background sources, within the redshift range $1.012\leq z \leq 5.793$ and randomly selected from the \ca\ catalog, to their multiple images on the lens plane. Both Cluster-2 and Cluster-3 amount to 69 magnified multiple images, whose positions are used as constraints for the lens model optimizations. As in Cluster-1, we still assume an isotropic statistical error of $0.2\arcsec$ on multiple image positions, but an additional systematic uncertainty on the position of multiple images and cluster member galaxies is considered in the Cluster-2, and Cluster-3 mocks. To do so, we assign random Gaussian displacements to the ``true'' positions, with a dispersion of $0.01\arcsec$ in both directions. These are comparable to the residual uncertainties from the HST-like quality of imaging data. 

In Fig.\,\ref{fig:DH}, we plot the total projected mass distribution of Cluster-2 and Cluster-3. Cluster member galaxies with measured velocity dispersions are encircled in green, while we mark the position of the ``observed'' multiple images using cyan crosses. The spatial distribution of the low-mass subhalo population included in the Cluster-3 mock is shown using magenta data-points on the right panel of the figure.

In Tab.\,\ref{table:model_ch}, we report a summary of the main proprieties of Cluster-1,2,3 mocks, such as: the number of galaxies with and without a measured velocity dispersion ($N_m^{gal}$ and $N^{gal}$), the number of low-mass subhalos ($N^{DH}$), the statistical and systematic errors on galaxy and multiple images positions ($\Delta x^{gal})_{sys}$, ($\Delta x^{im})_{st}$, ($\Delta x^{im})_{sys}$, and the number of observed multiple images ($N_{tot}^{im}$). 

\begin{figure}[t!] 
	\centering
	\includegraphics[width=9cm]{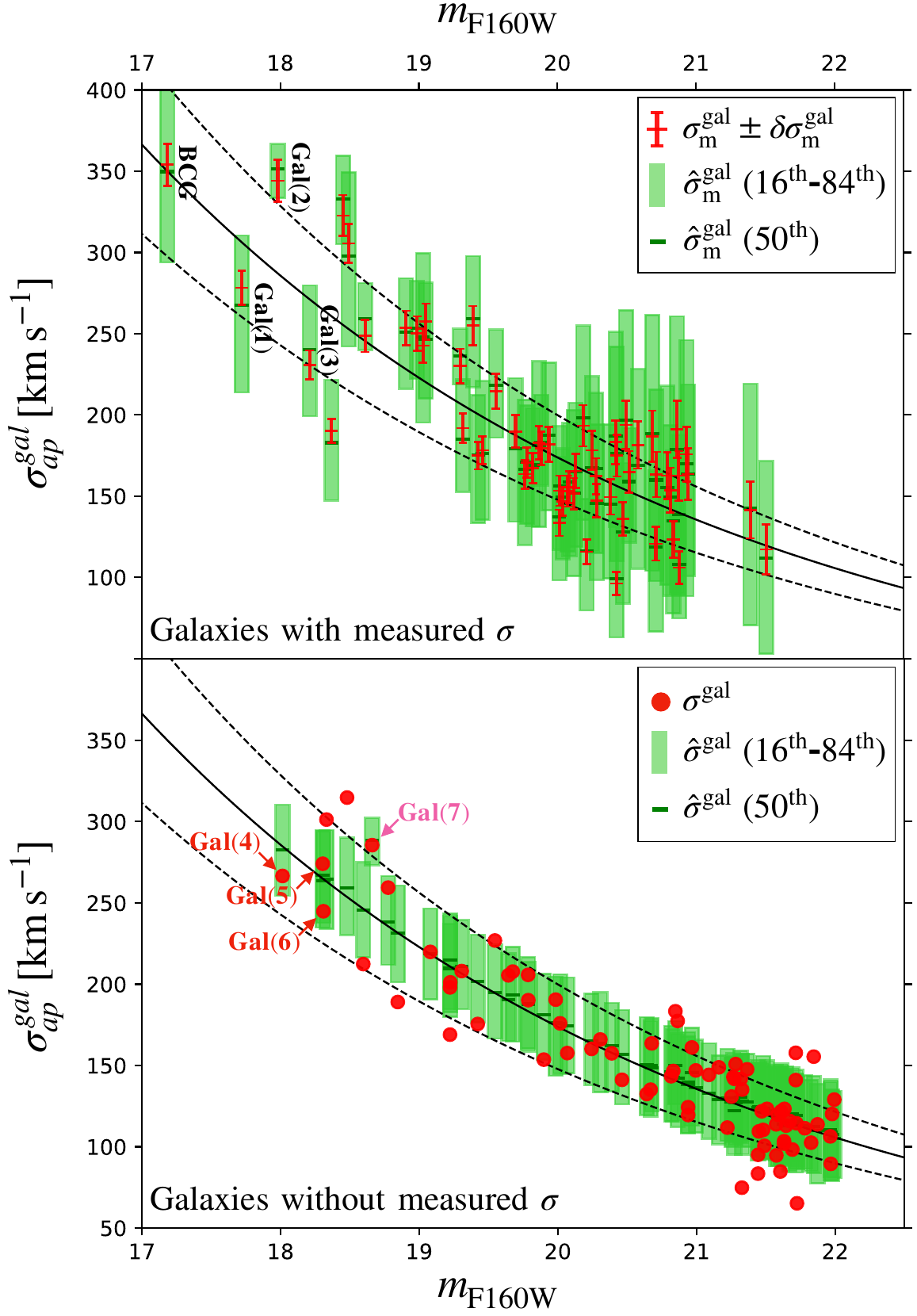}
	\caption{Velocity dispersions of the Cluster-1 member galaxies as a function of their magnitudes. In the mock cluster, we assume for the galaxy velocity dispersion a 15\% of Gaussian scatter (black dashed lines) around the scaling relation plotted as a solid black line. In the top panel, the ``measured'' velocity dispersions with their errors are plotted in red. Similarly, the ``true'' velocity dispersions of the galaxies without a measured velocity dispersion are marked with red dots in the bottom panel. The green rectangles in the plots are bounded by the 16th and 84th percentiles of the marginalized posterior distribution for each galaxy, while the 50th percentiles are the small green bars. We label in black and red the most luminous galaxies with and without ``measured'' velocity dispersion, respectively. We label in magenta the cluster member $Gal(7)$ forming the galaxy-scale strong lensing system displayed in the cut-out of Fig.\,\ref{fig:mock_cluster}.}
	\label{fig:green_plot}
\end{figure}

\begin{figure}[ht!] 
	\centering
	\includegraphics[width=9cm]{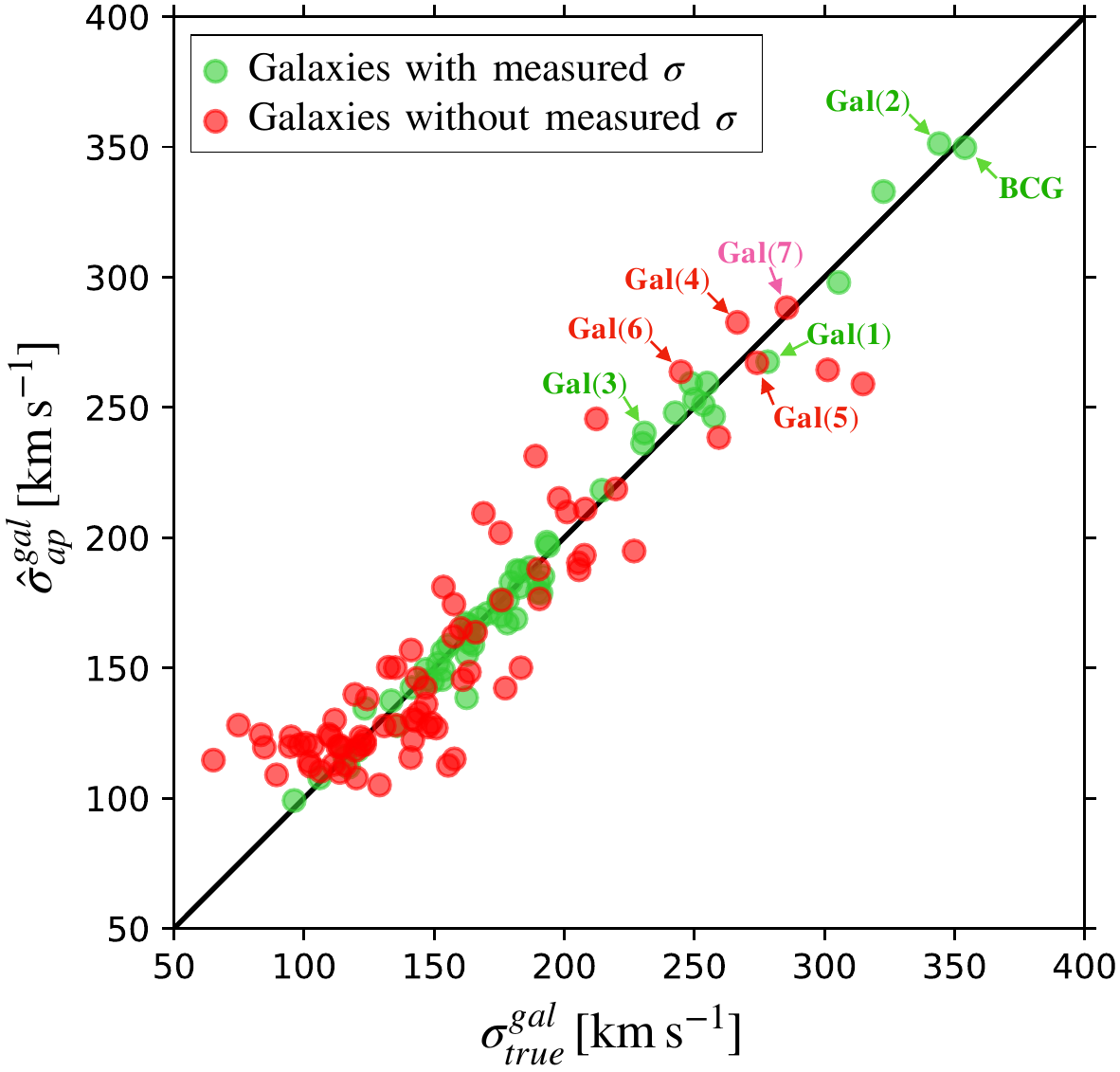}
	\caption{Velocity dispersions of cluster member galaxies from the \textsc{BayesLens} lensing model of Cluster-1 ($y\mbox{-}$axis) versus their ``true'' values in the mock ($x\mbox{-}$axis). The galaxies with and without a ``measured'' velocity dispersion are in green and red, respectively. In magenta the cluster member $Gal(7)$ responsible for the creation of the galaxy-scale strong lensing system displayed in the cut-out of Fig.\,\ref{fig:mock_cluster}.}
	\label{fig:velocities}
\end{figure}

\subsection{\textsc{BayesLens} on the mock clusters}

In this subsection, we describe the parameter ranges adopted in \textsc{BayesLens} to probe the mass distribution of the Cluster-1, Cluster-2, and Cluster-3 mocks.
The cluster-scale component of all three lens models is parametrized using a single PIEMD profile. The $x$, $y$ coordinates of its center can vary within flat priors, 3\arcsec\ wide, centered on the BCG position, while for the ellipticity $\hat{e}^h$, position angle $\hat{\theta}^{h}$, and fiducial velocity dispersion $\hat{\sigma}_{LT}^h$, we adopt uniform priors inside the following intervals respectively: 0.0\,-\,0.9, 5\,-\,$35^\circ$, and 700\,-\,1300\,\vel. Finally, for the PIEMD core radius $\hat{r}_{core}^h$, we assume a uniform prior between 1\arcsec and 7\arcsec in the Cluster-1 lens model and between 3\arcsec and 15\arcsec for Cluster-2 and Cluster-3.
The subhalo component of lens models is parametrized using the three scaling relations in \Eq\ref{eq.: Scaling_relation_sigma} normalized at the BCG luminosity. In particular, we fix $\hat{r}_{core}^{ref}=0.01\arcsec$ and $\hat{\beta}_{core}=0.5$, while the following flat priors are assumed on the others scaling relation parameters: Cluster-1 ($0.20 \leq \hat{\alpha} \leq 0.34$, $250\,$\vel $\leq \hat{\sigma}^{ref} \leq 450\,$\vel, $1\arcsec \leq \hat{r}_{cut}^{ref} \leq 11\arcsec$); Cluster-2, and Cluster-3 ($0.21 \leq \hat{\alpha} \leq 0.35$, $150\,$\vel $\leq \hat{\sigma}^{ref} \leq 350\,$\vel, $1\arcsec \leq \hat{r}_{cut}^{ref} \leq 9\arcsec$). Note that the slope of the $r_{cut}\mbox{-}mag$ scaling relation, $\hat{\beta}_{cut}$, is not a lens model free-parameter since its value is derived from $\hat{\alpha}$ through \Eq\ref{eq.: slopes} with $\gamma=0.2$. Differently from \textsc{LensTool}, in our code, the $\sigma\mbox{-}mag$ cluster member scaling relation refers to measured (line-of-sight) stellar velocity dispersions averaged within apertures of radius $R=0.8\arcsec$.

The \textsc{BayesLens} model optimization is performed on the lens plane using the positions of the multiple images inside the simulated catalog and the 58 ``measured'' stellar velocity dispersions of the cluster galaxies. These velocities are used on the one hand to determine the best $\sigma\mbox{-}mag$ scaling relation parameters, and on the other, to derive a Gaussian prior for each measured galaxy. Thanks to these priors, the lens model tends to prefer solutions with cluster members velocities that coincide with the measured values (see \Eq\ref{eq.: prior_measured_galaxies}), unless the lensing data require them to deviate. The presence of a non-zero scatter also enables the models to fairly sample the whole parameter space, thereby avoiding the underestimation of systematics.

In running \textsc{BayesLens} on the Cluster-1 mock, we leave all galaxies free to scatter around the fitted $\sigma\mbox{-}mag$ scaling relation, while in Cluster-2 and Cluster-3 lens models, only the 58 galaxies with a measured velocity dispersion, and the 22 most luminous unmeasured galaxies are left free to vary. The number of free-to-scatter galaxies is one of the inputs of our code, and in general, a lower number of free galaxies sensibly speed-up the optimization of the lens model.

The total posterior probability distribution ($p_{tot}$) of the Cluster-1 lens model is sampled using 298 walkers, while we use 500 walkers for the other two models. 
The initial walker positions are randomly distributed within the flat priors of the cluster-scale parameters ($\phi^h$), but they are initialized inside a Gaussian hyper-sphere, around an estimated maximum of the likelihood (in $p_{sr},p_{mg},p_g$), in the subspace of the other free-parameters. 
In the following, we derive the marginalized posterior distribution of model free-parameters by flattening the final Monte Carlo Markov chain (MCMC) of the walkers after removing a sufficiently large burn-in phase. Moreover, to determine the model-predicted multiple image positions and the r.m.s. displacement, $\Delta^{tot}_{rms}$, we use the best-fit \textsc{BayesLens} model. This best-fit model is defined as the set of free-parameter values that maximize the total posterior $p_{tot}$.

\begin{figure}[t!] 
	\centering
	\includegraphics[width=9cm]{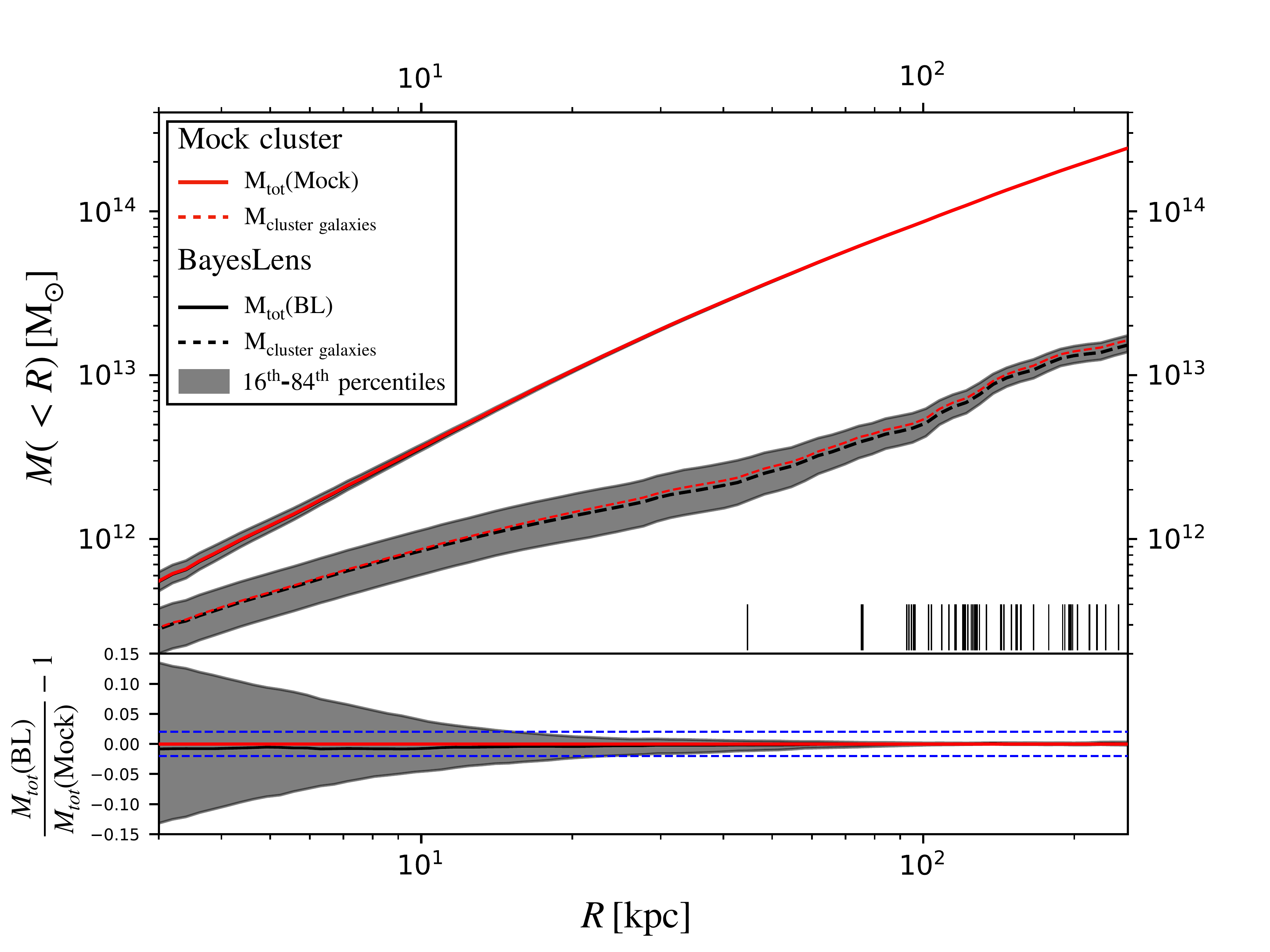}
	\caption{\textit{Top}: Cumulative projected total mass profiles for the Custer-1 mock as a function of the projected distance from the center of the BCG. The red line corresponds to the true mass profile. The solid black line is obtained from the best-fit \textsc{BayesLens} model. The projected distances of the multiple images from the BCG center are shown as vertical black lines. \textit{Bottom}: Relative variation of the \textsc{BayesLens} total mass profile with respect to the true mass profile of the mock cluster. The dashed blue line in the bottom panel marks a 2\% relative difference between the mock and model mass profiles. In both panels, the gray shaded stripes delimit the 16th-84th percentiles of the mass profiles with parameters drawn from the posterior.}
	\label{fig:mass}
\end{figure}

\section{Results and discussion}
\label{sec:results}

In this section, we describe the main results obtained from an application of \textsc{BayesLens} on the three simulated clusters, namely Cluster-1, Cluster-2, and Cluster-3.

\subsection{BayesLens results for Cluster-1}

In Fig.\,\ref{fig:degeneracies}, we show in black the marginalized \textsc{BayesLens} posterior distributions for the free-parameters of the Cluster-1 lens model. The ``true'' parameter values assumed in the mock cluster are marked with solid red lines.  
Panels $\boldsymbol{a}$ and $\boldsymbol{b}$ show the posterior distributions of the scaling relation and cluster-scale halo parameters, respectively. In panel $\boldsymbol{c}$, we plot the marginalized posterior distributions for the stellar velocity dispersions of the four brightest ``measured'' cluster galaxies: $\hat{\sigma}_{m}^{\mathrm{BCG}}$, $\hat{\sigma}_{m}^{\mathrm{Gal(1)}}$, $\hat{\sigma}_{m}^{\mathrm{Gal(2)}}$ and $\hat{\sigma}_{m}^{\mathrm{Gal(3)}}$. Similarly, panel $\boldsymbol{d}$ shows the velocity dispersions of the three brightest cluster galaxies without measured kinematics ($\hat{\sigma}^{\mathrm{Gal(4)}}$, $\hat{\sigma}^{\mathrm{Gal(5)}}$ and $\hat{\sigma}^{\mathrm{Gal(6)}}$) and of the lens galaxy,  $\hat{\sigma}^{\mathrm{Gal(7)}}$, forming the galaxy scale strong lensing system zoomed in Fig\,\ref{fig:mock_cluster}. In these last two panels, we plot also blue solid lines showing the value of galaxy stellar velocity dispersions, as predicted by the best-fit $\sigma^{gal}\mbox{-}m_{\mathrm{F160W}}^{gal}$ scaling relation. The ``true'' and \textsc{BayesLens} optimized values of the lens model free-parameters are reported in Tab.\,\ref{table:input_output}. The latter correspond to the medians of parameter marginalized posterior distributions, while the 16th and 84th percentiles are quoted as an error.
Thanks to the ``measured'' cluster members stellar velocity dispersions and to the observed multiple-image positions, \textsc{BayesLens} recovers, well within the $1\sigma$ uncertainty, all of the mock true hyperparameters of the scaling relations and the cluster-scale halo parameters.
To quantify the accuracy of \textsc{BayesLens} in determining the correct stellar velocity dispersions of the cluster members, we plot in Fig.\,\ref{fig:green_plot} the aperture average velocity dispersions within aperture of 0.8\arcsec radius,  $\sigma^{gal}$, as a function of the galaxy magnitudes, $m_{\mathrm{F160W}}$. The solid black lines correspond to the scaling relation, with the dashed lines marking the 15\% scatter in the mock. The (red) errorbars in the upper panel denote the ``measured'' velocity dispersions with their uncertainties, and the (green) rectangles mark the 16th-to-84th percentiles from the posterior distributions. All of the galaxies lie on their measured velocity dispersions, and the zero-scatter solution still exists within the sampled parameter space. This shows that the loose prior on ``measured'' galaxies still allows the models to generalize the zero-scatter ansatz. The bottom panel shows the mock and posterior dispersions for the ``unmeasured'' galaxies. Most of them are compatible with their true values (red dots) within the uncertainties. Most notably, one galaxy-scale strong lensing system (marked as Gal(7), see Fig.\,\ref{fig:mock_cluster} closeup) departs automatically form the zero-scatter {best-fit} scaling relation and has a dispersion that is very well constrained by the lensing data. The compatibility of mock and posterior dispersions for all galaxies (within the intrinsic scatter) is displayed in Fig.\,\ref{fig:velocities}, showing that there is no appreciable systematic mismatch trend between the ``true'' and inferred dispersions.

Fig.\,\ref{fig:mass} shows the cumulative mass profile of the whole cluster, projected along the line of sight, for the mock and the best-fit (maximum a posteriori) model. Even though the total mass of the mock cluster is redistributed in a slightly different way between the cluster halo and the member galaxies, its value at large radii is accurately recovered by the lens model. At the distances where most of the multiple images form, yielding more constraints to the lens model, the model predicts a correct value for the mass profiles within 1\%. This is simply because the projected mass $M(<R_{E})=\pi\Sigma_{cr}R_{E}^{2}$ within the Einstein radius of a lens does not depend on its density profile.

To relate the goodness of fit of the models to a commonly used benchmark, we examine the r.m.s. offset of ``observed'' vs inferred positions of the multiple images ($\Delta_{rms}^{tot}$). Our best-fit hierarchical model for Cluster-1 can reproduce all images with $\Delta_{rms}^{tot}=0.08\arcsec$, while the lens model total chi-square is $\chi^{2}_{tot}=8.66$ (see Tab.\,\ref{table:model_ch}). This non-zero $\Delta_{rms}^{tot}$ shows that our hierarchical model does not over-fit while also reproducing all observables well within their uncertainties. 

The robustness against over-fitting can be understood in two ways. First, if we re-run the models on a mock with $0.05\arcsec$ statistical uncertainties on the image positions, the maximum-a-posteriori parameters do not change, and the r.m.s. offset is still $\approx0.08\arcsec$, while only the confidence intervals in the inferred parameters are shrunk. This is because all image-positions in this first mock are kept on their true locations, and the statistical uncertainties are simply a tolerance parameter that enables a smooth exploration of parameter space. Second, from the posterior we see that only about ten galaxies deviate by more than $1\sigma$ from the backbone of the scaling relation; if we re-run a \textsc{LensTool} model with only these galaxies free, and all other galaxies on the zero-scatter scaling relation, the $\chi^{2}$ improves by $\approx33$, that is three times the change in degrees of freedom.

\begin{figure*}[h!] 
	\centering
	\includegraphics[width=\textwidth]{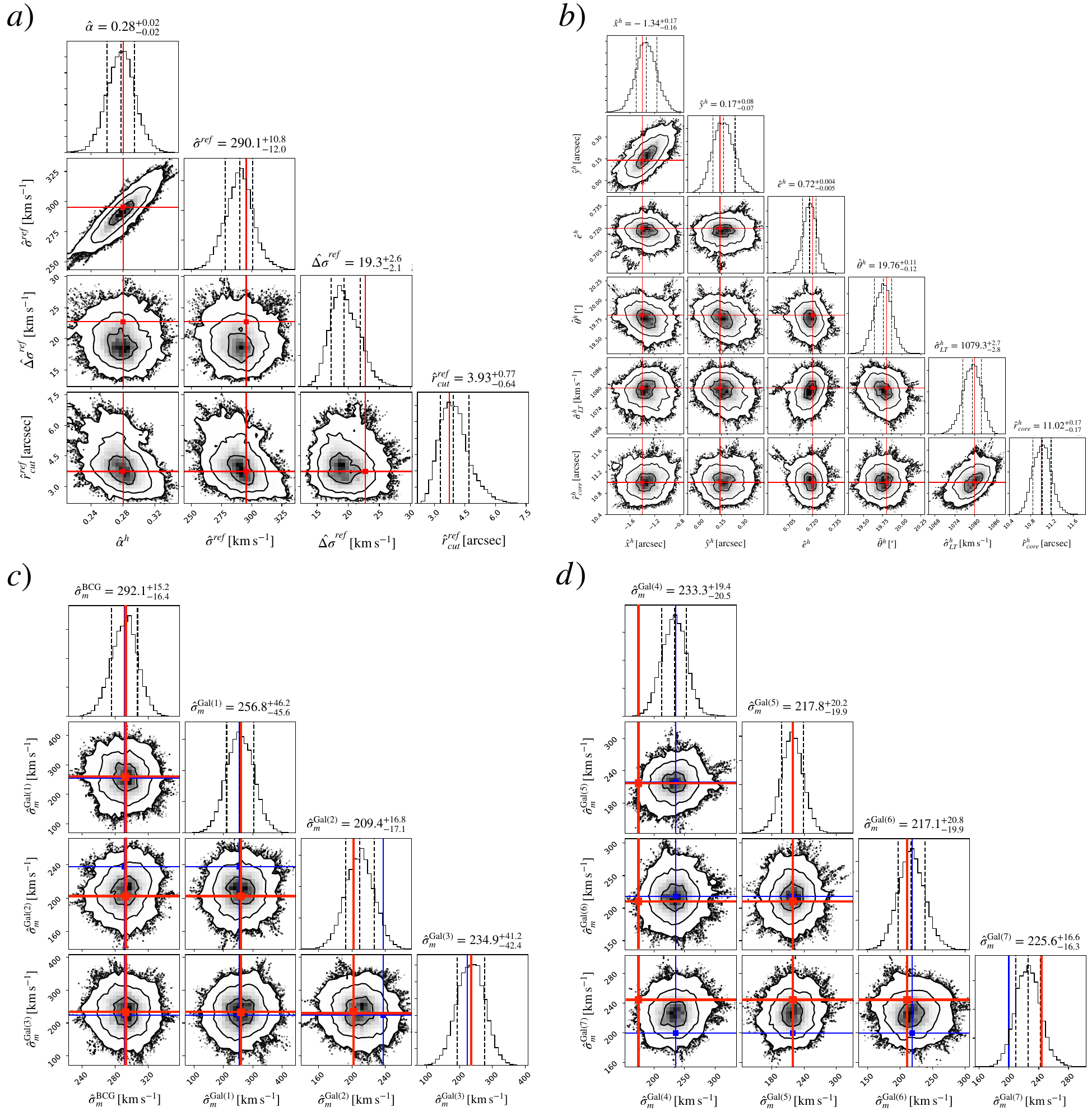}
	\caption{{Same as Fig.\,\ref{fig:degeneracies} but for the Cluster-3 mock.}}
	\label{fig:DH_deg}
\end{figure*}

\subsection{BayesLens results for Cluster-2 and Cluster-3}

A comparison between ``true'' values of the Cluster-2 and Cluster-3 parameters with \textsc{BayesLens} results is in Tab.\,\ref{table:input_output}. Note that in this table we are showing the same set of free-parameters for the three mocks, however, their ``true'' values are different in the Cluster-1 and Cluster-2,3 mocks. 

Despite the increasing complexity of the last two mock clusters, \textsc{BayesLens} recovers well within $1\sigma$ all cluster-scale halo and scaling relations parameters. Moreover, it finds similar values for the stellar, aperture-average, velocity dispersions of measured and unmeasured cluster galaxies in the two mocks.

In Fig\,\ref{fig:DH_deg}, we plot marginalized posterior distributions, obtained from the Cluster-3 MCMC chains, for the same set of model parameters reported in Tab.\,\ref{table:input_output}. 
The figure shows that the four measured galaxies BCG, Gal(1), Gal(2), Gal(3), and the two unmeasured galaxies Gal(5), Gal(6) have marginalized posterior distributions that pick very closed to the ``true'' mock stellar velocity dispersions (red vertical lines). Conversely, we observed large deviations from ``true'' values for Gal(7) and even more for Gal(4). The reason for these deviations is at the basis of the hierarchical approach implemented in \textsc{BayesLens}, and it is due to the interplay between the $p_{g}$ and $p_{im}$ terms of the total posterior in \Eq\ref{eq.: posterior_tot}. In fact, the unmeasured galaxy Gal(4) has marginalized posterior distribution centered on the stellar velocity dispersion value predicted by the best-fit $\sigma^{gal}\mbox{-}m_{\mathrm{F160W}}$ scaling relation (marked with a blue line in Fig\,\ref{fig:DH_deg}). In the absence of constraints from the position of the observed multiple images, the $p_{g}$ term encourages a zero-scatter lens model solution with all unmeasured galaxy velocity dispersions that lie on the best-fit scaling relation. 
Similarly, the displacement of the $\hat{\sigma}^{\mathrm{Gal(7)}}$ posterior distribution from the scaling relation toward the ``true'' mock value is possible because the increase of the $p_{im}$ term (due to a lower r.m.s displacement of the images close to Gal(7)) overcome the decrease of $p_{g}$ that penalizes scattered solutions.

The best-fit Cluster-2 lens model reproduces the position of multiple images with a $\Delta_{rms}^{tot}=0.09\arcsec$, while the total chi-square is $\chi^2_{tot}=12.53$. As displayed in Tab.\,\ref{table:model_ch}, the addition of the low-mass subhalo population in the Cluster-3 mock produces a small increase, equal to 1.22, of the lens model total chi-square, however, the final $\Delta_{rms}^{tot}$ of multiple images remains essentially unchanged.

\subsection{Testing against main halo mismatch}
\label{sec:halo_mismatch}

Different mass models, related through unobservable source-position transformations, may be able to reproduce the same image positions, even though these degeneracies are more severe for systems with less symmetry \citep{Schneider14,Unruh17}. This occurrence is common to all parametric lensing codes. Here, we check whether \textsc{BayesLens} uses the non-zero scatter to over-fit models that are different from the mocks.

To test whether this happens, we also considered one last simulation whose main halo has a Navarro-Frenk-White (hereafter NFW) density profile, while the \textsc{BayesLens} model still adopt a PIEMD to parameterize the main cluster halo. The cluster-scale NFW halo has a total mass $M_{200c}=1.59\times10^{15}\,\mathrm{M_{\odot}}$ within a radius $r_{200c}=2.06\,\mathrm{Mpc}$. These values correspond to those measured by \cite{Umetsu_2014}, using weak lensing techniques, for the cluster MACS J1206.2$-$0847. Assuming a concentration $c_{200c}=3.5$, we derive a scale radius $r_s=588.6\,\mathrm{kpc}$. The center position of the NFW profile has an offset of 1.4\arcsec\ from the cluster BCG, an ellipticity of 0.4, and a position angle of $19.8^\circ$ counterclockwise from west. The subhalo component of the cluster and the position of the lensed sources are assumed the same as the Cluster-2 simulation. The lens model for this cluster is constrained by 69 ``observed'' multiple images, and systematic isotropic uncertainties of $0.01\arcsec$ are assumed on galaxies and image positions (see Cluster-2 model for more details).

The r.m.s. displacement of the best-fit \textsc{BayesLens} model is 0.40\arcsec, that is about 0.32\arcsec\ higher than the $\approx0.08$\arcsec\ of a PIEMD model on a PIEMD mock. Some of the observed multiple images are not produced by the best-fit model, whereas the PIEMD models on PIEMD mocks (Cluster-1,2,3) were all able to produce the whole set of images. We re-run \textsc{BayesLens} while also penalizing models for each multiple image that cannot be produced by the model\footnote{We add -100 to the log-likelihood for each multiple image that is not produced by the lens model.}. In this case, the best-fit r.m.s. displacement is just 0.35\arcsec, and all the images are well reproduced by the model.

The slightly higher r.m.s. displacement, as well as the new issues in producing all multiple images, indicate that a main-halo mismatch may be diagnosed when modeling real clusters. \textsc{BayesLens} does not seem to over-fit the most evident systematics, despite higher freedom in the cluster member galaxies.

\begin{figure*}[t!] 
	\centering
	\includegraphics[width=13cm]{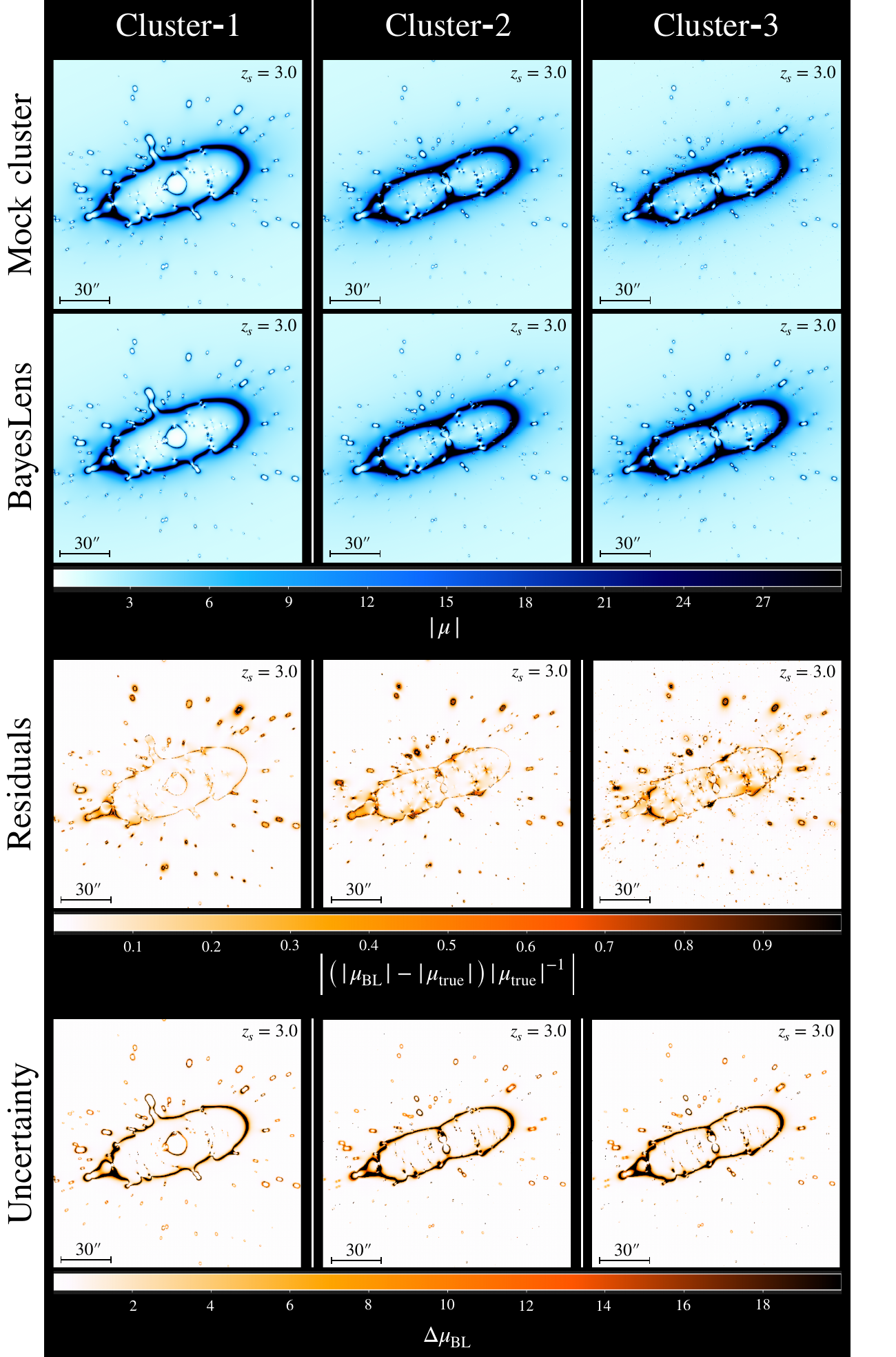}
	\caption{Maps of absolute magnification for the three simulated clusters computed for a source redshift $z_s=3.0$. In the first line, we plot magnification maps for the ``true'' Cluster-1, Cluster-2, and Cluster-3 mocks. Maps in the second line are obtained from BayesLens best-fit lens models of the clusters, while in the third line, we show normalized residuals between BayesLens and true magnifications: $\left| |\mu_\mathrm{BL}|-|\mu_\mathrm{true}|/|\mu_\mathrm{true}|\right|$. In the last line, we plot absolute uncertainties associated with \textsc{BayesLens} magnification maps. These are computed considering 100 realizations of the lens models by randomly extracting 100 parameter samples from the MCMC chains and taking half of the difference between the 84th and 16th percentiles of magnification distributions in each pixel.}
	\label{fig:magnification}
\end{figure*}

\section{Conclusions}
\label{sec:conclusions}

We have shown that, despite the (nominally) large number of degrees of freedom in cluster lensing, the exploration of flexible models is feasible thanks to tractable, hierarchical Bayesian inference. Unlike conventional models, where galaxies are placed on a razor-thin scaling relation, and some are ``freed'' ad hoc, we populate a scaling relation with non-zero scatter, with hyperparameters that are inferred directly from the lensing constraints and (when available) stellar kinematic data. Our tests on a realistic, albeit simplified, mock cluster (namely Cluster-1) show that the improvement over conventional models may be significant: the velocity dispersions of all the galaxies of our mock cluster are reliably recovered, as are the scaling relation hyperparameters, and the r.m.s. displacement of multiple images between measurements and model predictions can decrease appreciably (in the simplest mocks, from 0.40\arcsec\ to 0.08\arcsec) unless there are other major sources of systematics. As shown by the tests in \Sec\ref{sec:halo_mismatch}, the impact of the non-zero scatter is mainly in the deflections from the cluster member galaxies and cannot correct for a mismatch between the `true' density profile of the main cluster halo and the one used in the models. This also has some practical implications when modeling real-life clusters: if the r.m.s. displacements do not change appreciably from zero-scatter to hierarchical models, this would indicate that other major sources of systematics must be analyzed (e.g., the density profile of the main halo, or major mass contributions along the line of sight). We also remark that these models do not make any claim on the internal mass content of the galaxies themselves, but just on the fact that they do not necessarily have simple PIEMD profiles, that there is a real and measured scatter in their properties, and that the information from lensing is complementary to that from the measured stellar kinematics (i.e., the hyperparameters are free to differ from the ones from stellar kinematics).

We have also tested the robustness of the hierarchical models against two systematic effects: dark subhalos and small systematic offsets in the measured image and cluster member positions (Cluster-2 and Cluster-3 mocks). Within realistic regimes, compatible with current (HST) or upcoming (ELT) depth and image quality, the model performance on the mock cluster does not degrade appreciably.

A crucial consequence of our hierarchical approach is that it also resolves a ``false dichotomy'' between well-fitting cluster models and realistic population properties of galaxies. 
Zero-scatter models around a kinematic prior have realistic galaxy populations but a higher r.m.s. offset than models without any kinematic prior, whose scaling relations are significantly discrepant from the ``true'' input relation.
In zero-scatter models, there is typically a trade-off between the r.m.s. and recovering realistic galaxy populations \citep[see e.g.,][]{ber19}.
Our hierarchical models have realistic galaxy populations and produce a small r.m.s. offset (on our mock cluster). With respect to zero-scatter models, \textsc{BayesLens} automatically identifies the galaxy-scale systems where more freedom from the scaling relation backbone is needed in order to reproduce the positions of multiple images around those galaxies with high accuracy. This makes the hierarchical models robust against over-fitting since, in our mocks, only about ten galaxies deviate from the scaling relation backbone by more than 1$\sigma$ while the chi-square improves by $\Delta\chi^{2}\approx33.$

We should emphasize that this is a functional test on mocks, and additional effects may play a role in real-life systems, such as massive substructure, deviations from simple geometry of the main DM halo(s) and cluster members, and additional contributions along the line of sight. 
However, our hierarchical models can eliminate part of the systematic source of uncertainties and show that internal systematics can be controlled.

The freedom in the parameters of each cluster member galaxy, within the intrinsic scatter of their parent population, has multiple implications. First, the data themselves dictate which galaxy should be ``freed'' and deviate significantly from a baseline scaling relation. Second, intrinsic scatter is a hyperparameter that is left free in the modeling inference, and this allows for a direct determination of galaxy-population properties, whence accurate studies of galaxy formation and evolution. In particular, accurate mass functions of cluster member galaxies can be compared to predictions from cosmological simulations. Also, the freedom in cluster member parameters allows (in principle) for the quantification of environmental effects on galaxies at different locations across the cluster -- which up to now has been possible only on stacked weak-lensing measurements \citep{nie17}.

An important feature of our inference is that \textsc{BayesLens} is fully modular. Changing specific python functions within our code, it is possible, in theory, to allow for calls to any chosen parametric lensing software (besides \textsc{LensTool}, used here as the benchmark), as well as to implement different mass profiles. Moreover, different prescriptions to relate the stellar velocity dispersions and lensing parameters of cluster member galaxies can be used. The currently used scaling relations may be easily replaced with fundamental-plane relations (with free hyperparameters) so as to study the evolution of the fundamental plane across redshift and the environment. Its modular structure also enables the use of additional constraints from, for example, flux ratios or extended-source reconstruction. Samples of magnification maps may be produced simply through an other module, which may enable accurate studies of high-redshift galaxy populations.

This is not central to this first paper, so for the sake of brevity, in Fig.\,\ref{fig:magnification}, we show simply a comparison of magnification maps from the best-fit models of the three mock clusters (namely Cluster-1,2,3). The magnification maps are well reproduced, and as can be expected, the appreciable differences happen only across the critical lines and in high magnification regions ($|\mu|\gtrsim20$), where the posterior also predicts a wider variance in the magnification. We remark that these models are already an extension over the state of the art, and the availability of samples from the posterior allows one to evaluate magnification uncertainties at all points.

While our set of hyperparameters is currently limited to those of the scaling relations (intercepts, slopes, scatter widths), this can be in principle extended to cosmological parameters, enabling (fully self-consistent) cosmographic measurements.
Our code is released publicly\footnote{The GitHub link is proprietary until acceptance.} and documented at \url{https://github.com/pietrobergamini89/BayesLens}.

\begin{acknowledgements}
PB and GBC thank Piero Rosati, Claudio Grillo, Massimo Meneghetti and Amata Mercurio who have made this work possible. We thank A.~Sonnenfeld and M.~Lombardi for useful comments on a previous version of this paper. PB acknowledges financial support from PRIN-MIUR 2017WSCC32ASI and through the agreement ASI-INAF n. 2018-29-HH.0. AA was supported by a grant from VILLUM FONDEN (project number 16599). This project is partially funded by the Danish council for independent research under the project ``Fundamentals of Dark Matter Structures'', DFF--6108-00470. GBC acknowledge funding from the European Research Council through the Grant ID 681627-BUILDUP, the Max Planck Society for support through the Max Planck Research Group for S. H. Suyu and the academic support from the German Centre for Cosmological Lensing.
\end{acknowledgements}

\bibliography{bibliography}





\end{document}